\documentclass[journal]{IEEEtran}
\ifCLASSINFOpdf

\else

\fi

\usepackage{CJK}
\usepackage{algorithm}
\usepackage{algorithmic}
\usepackage{amsmath}
\usepackage{amssymb}
\usepackage{psfrag}
\usepackage{graphicx}
\usepackage{epstopdf}
\usepackage{multirow}
\usepackage{stfloats}
\usepackage{cite}
\usepackage{subcaption}
\usepackage{color}
\usepackage[normalem]{ulem}
\usepackage{arydshln}
\usepackage{enumerate}
\usepackage{bbm}

\newtheorem{theorem}{\bf{Theorem}}[section]
\newtheorem{lemma}{\bf{Lemma}}[section]

\newcommand{\bm}[1]{\mbox{\boldmath{$#1$}}}

\begin{document}

\title{Intelligent Reflecting Surface: A Programmable Wireless Environment for Physical Layer Security}
\author{ Jie Chen, \IEEEmembership{Student Member, IEEE}, Ying-Chang Liang, \IEEEmembership{Fellow, IEEE}, Yiyang Pei, \IEEEmembership{Member, IEEE},\\ and Huayan Guo, \IEEEmembership{Member, IEEE}

\thanks{This work was supported in part by the National Natural Science Foundation of China under Grants 61631005, U1801261, and 61571100.
J. Chen and H. Guo are with the National Key Laboratory of Science and Technology on Communications, and Center for Intelligent Networking and Communications (CINC), University of Electronic Science and Technology of China (UESTC), Chengdu 611731, China (e-mails: chenjie.ay@gmail.com and   guohuayan@uestc.edu.cn). Y.-C. Liang is with the Center for Intelligent Networking and Communications (CINC), University of Electronic Science and Technology of China (UESTC), Chengdu 611731, China (e-mail:liangyc@ieee.org). Y. Pei is with the Singapore Institute of Technology, Infocomm Technology Cluster, Singapore 138683 (e-mail: yiyang.pei@singaporetech.edu.sg). }
}

\maketitle

\begin{abstract}
In this paper, we introduce an {\emph{intelligent reflecting surface}} (IRS) to provide a programmable wireless environment for physical layer security. By adjusting the reflecting coefficients, the IRS can change the attenuation and scattering of the incident electromagnetic wave so that it can propagate in a desired way toward the intended receiver. Specifically, we consider a downlink {\emph {multiple-input single-output}} (MISO) broadcast system where the {\emph{base station}} (BS) transmits independent data streams to multiple legitimate receivers and keeps them secret from multiple eavesdroppers.
 By jointly optimizing the beamformers at the BS and reflecting coefficients at the IRS, we formulate a minimum-secrecy-rate maximization problem under various practical constraints on the reflecting coefficients. The constraints capture the scenarios of both continuous and discrete reflecting coefficients of the reflecting elements.
Due to the non-convexity of the formulated problem, we propose an efficient algorithm based on the alternating optimization and the path-following algorithm to solve it in an iterative manner.
Besides, we show that the proposed algorithm can converge to a local (global) optimum. Furthermore, we develop two suboptimal algorithms with some forms of closed-form solutions to reduce the computational complexity. Finally, the simulation results validate the advantages of the introduced IRS and the effectiveness of the proposed algorithms.
\end{abstract}

\IEEEpeerreviewmaketitle

\section{Introduction}

A variety of wireless technologies have been proposed to enhance the spectrum- and energy-efficiency due to the tremendous growth in the number of communication devices, such as {\emph {multiple-input multiple-output}} (MIMO)\cite{rusek2013scaling}, cooperative communications \cite{nosratinia2004cooperative}, {\emph{cognitive radio}} (CR) \cite{Liang2011Cognitive} and so on.
However, these techniques only focus on the signal processing at the transceiver to adapt the changes of the wireless environment, but cannot eliminate  the negative effects caused by the uncontrollable electromagnetic wave propagation environment \cite{liaskos2018new,yang2016programmable}.

Recently, {\emph{intelligent reflecting surface} (IRS) has been proposed as a promising technique due to its capability to achieve high spectrum-/energy-efficiency through controlling the wireless propagation environment \cite{cui2014coding}. Specifically, IRS is a uniform planar array consisting of a large number of composite material elements, each of which can adjust the reflecting coefficients (i.e., phase or amplitude) of  the incident electromagnetic wave and reflect it passively.
Hence, by smartly adjusting the reflecting coefficients with a preprogrammed controller, the IRS can change the attenuation and scattering of the incident electromagnetic wave so that it can propagate in the desired way before reaching the intended receiver, which is called as programmable and controllable wireless environment. This also inspires us to design the communication systems by jointly considering the signal processing at the transceiver and the optimization of the electromagnetic wave propagation in the wireless environment.

Compared with the existing related techniques, i.e., traditional reflecting surfaces \cite{ford1984electromagnetic}, {\emph{amplify-and-forward}} (AF) relay \cite{zhang2009optimal}, active intelligent surface \cite{hu2018beyond}, and backscatter communication \cite{yang2018modulation,li2018adaptive,Guo2019Exploiting}, IRS has the following advantages \cite{qingqing2019towards,wu2018intelligent1}.
Firstly, IRS can reconfigure the reflecting coefficients in real time thanks to the recent breakthrough on {\emph{micro-electrical-mechanical systems}} (MEMS) and composite material \cite{cui2014coding,yang2016programmable} while the traditional reflecting surface only has fixed reflecting coefficients. Secondly, IRS is a green and energy-efficient technique which reflects the incident signal passively without additional energy consumption while the AF relay and the active intelligent surface require active {\emph{radio frequency}} (RF) components. Thirdly, although both the IRS and the backscatter communication make use of passive communications, IRS can be equipped with a large number of reflecting elements while backscatter devices are usually equipped with a single/few antenna(s) due to the limitations of complexity and cost \cite{Long2019FullDuplex}. Besides,   IRS only attempts to assist the transmission of the signals between the intended transmitter and receiver pair with no intention for its own information transmission while backscatter communication needs to support the information transmission of the backscatter device \cite{yang2018cooperative,Zhang2019Constellation}.

Due to the significant advantages,  IRS has been introduced into various wireless communication systems.
Specifically, \cite{Ruizhang2018Sufface,han2018large,tan2018enabling,tan2016increasing,wu2018beamformingOptimization} consider a downlink single user {\emph {multiple-input single-output}} (MISO) system assisted by the IRS. In \cite{Ruizhang2018Sufface}, both centralized and distributed algorithms were developed to maximize the {\emph{signal-to-noise ratio}} (SNR) of the desired signals considering perfect {\emph{channel state information}} (CSI). Then, in \cite{han2018large}, the effect of the reflecting coefficients on the ergodic capacity was investigated by considering statistical CSI. Moreover, since achieving continuous reflecting coefficients  on the reflecting elements is costly in practice due to the hardware limitation, the SNR maximization problem and transmitter power minimization problem were studied in  \cite{tan2018enabling,tan2016increasing,wu2018beamformingOptimization,wu2019beamforming} by considering discrete reflecting coefficients on the reflecting elements.
As for a downlink multi-user MISO system \cite{huang2018large,huang2018energy,nadeem2019largearxiv},  the spectrum-/energy-efficiency problem under the individual {\emph{signal-to-interference-plus-noise ratio}} (SINR) constraints was investigated in \cite{huang2018large} and \cite{huang2018energy} considering continuous or discrete  reflecting coefficients on the reflecting elements. In addition, the minimum-SINR maximization problem was formulated in \cite{nadeem2019largearxiv} by considering the two cases where the channel matrix between the transmitter and the IRS is of rank-one and of full-rank.

Furthermore, physical layer security is a fundamental issue in wireless communications \cite{pei2011secure}. The basic wiretap channel introduced by Wyner \cite{wyner1975wire} consists of one transmitter, one legitimate receiver, and one eavesdropper. Then, the basic wiretap channel has been extended to broadcast channels \cite{csiszar1978broadcast}, Gaussian channels \cite{leung1978gaussian}, compound wiretap channels \cite{liang2009compound}, and so on.
It is worth noting that, in order to ensure secret communications, the transmission rate in the wiretap channel should be lower than the secrecy capacity of the channel. Thus,
 MIMO beamforming techniques were further introduced to improve the secrecy capacity (improving SNR of legitimate receivers and  suppressing SNR of eavesdroppers) \cite{pei2010secure,cumanan2014secrecy,LiuSecrecy2014,nasir2017secrecy}. Specifically, both power minimization and  secrecy rate maximization were studied in \cite{cumanan2014secrecy} in a single user/eavesdropper MIMO systems considering both perfect and imperfect CSI. Then, the minimum-secrecy-rate of a single-cell multi-user MISO system was studied in \cite{LiuSecrecy2014} with a minimum harvested energy constraint, and it was further extended to a multi-cell  network in \cite{nasir2017secrecy}.

However, consider the special case when the legitimate receivers and the eavesdroppers are in the same directions to the transmitter. In this case, the channel responses of the legitimate receivers will be highly correlated with those of the eavesdroppers. The beamformers proposed in \cite{pei2010secure,cumanan2014secrecy,LiuSecrecy2014,nasir2017secrecy} to maximize the SNR of legitimate receivers will also maximize the SNR of eavesdroppers. Hence,  it is intractable to guarantee the secret communications with the use of beamforming only at the
transceivers. Hence, we want to explore the use of the IRS to provide additional communication links so as to increase the SNR at the legitimate receivers while suppressing the SNR
at the eavesdroppers. Hopefully, this will create an effect as if the confidential data streams can bypass the eavesdroppers and reach the legitimate receivers, as shown in Fig. \ref{fig:fig1}, and thus the secrecy rate will be improved.

Motivated by the above reasons, in this paper, we study a programmable wireless environment for physical layer security to achieve high-efficiency secret communication. Specifically, we consider a downlink MISO broadcast system where the {\emph{base station}} (BS) transmits multiple independent confidential data streams to each legitimate receivers and keeps them secret from the eavesdroppers through the assistance of the IRS.
The contributions of the paper are summarized as follows:
\begin{itemize}
\item To the best of our knowledge, this is the first work to explore the use of the IRS to enhance the physical layer secret communication. Particularly, we jointly optimize the beamformers at the BS and the reflecting coefficients at the IRS to maximize the minimum-secrecy-rate under various practical constraints on the reflection coefficients. The constraints capture both the continuous and discrete reflecting coefficients of the reflecting elements on the IRS. However, the objective function is not jointly concave with respect to both the beamformers and the reflecting coefficients, and even worse, they are coupled together. Hence, the formulated problem is non-convex, which is hard to solve and may require high complexity to obtain the optimal solutions.
\item We solve the formulated problem efficiently in an iterative manner by developing alternating optimization based path-following algorithm \cite{zaslavskiy2009path,sheng2018power}. Specifically, we use the path-following algorithm to handle the non-concavity of the objective function and apply the alternating optimization to deal with the coupled optimization variables.  Besides, we prove that  the proposed algorithm is guaranteed to converge to a local (global) optimum and the corresponding solution will converge to a {\emph{Karush-Kuhn-Tucker}} (KKT)  point finally.
\item To further reduce the computational complexity, we develop two suboptimal algorithms to solve the formulated problem for two cases. For the first case with one legitimate receiver and one eavesdropper, we develop an alternating optimization method to solve the formulated problem in an iterative manner, but in each iteration we provide the closed-form solutions, which leads the algorithm to be low complexity. For the second case with multiple legitimate receivers and eavesdroppers, we develop a heuristic closed-form solution based on {\emph{ zero-forcing}} (ZF) beamforming, which further reduces the computational complexity.

\item Finally, the simulation results validate the advantages of the introduced IRS and also show the effectiveness of the proposed algorithms.
\end{itemize}
\begin{figure}
    \centering
        \includegraphics[width = 0.4\textwidth]{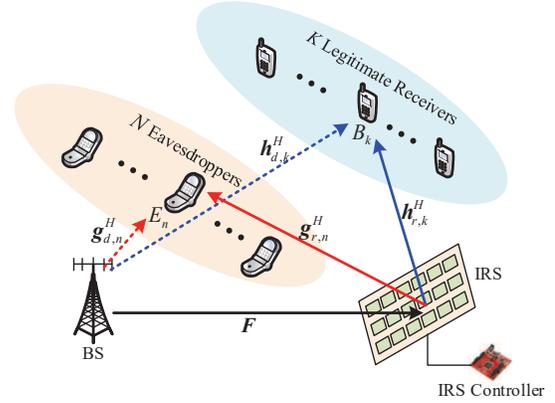}
    \caption{A programmable downlink MISO broadcast system with one IRS and multiple eavesdroppers.}
    \label{fig:fig1}
\end{figure}

The rest of this paper is organized as follows:
Section~\ref{sec:System Model} introduces the system model of the downlink MISO broadcast system with multiple eavesdroppers.
Section \ref{secProblemStatement} formulates the minimum-secrecy-rate maximization problem.
Section~\ref{secProblemSolution} develops an efficient algorithm to solve the formulated problem and Section~\ref{secLowcomplexity} provides two low-complexity suboptimal algorithms to solve it in two cases, respectively. Section~\ref{secSimulation} shows the simulation results to evaluate the performances of the proposed algorithms.
Finally, Section~\ref{secConclusion} concludes the paper.

The notations used in this paper are listed as follows. The scalar, vector, and matrix are lowercase, bold lowercase, and bold uppercase, i.e., $a$, $ {\bf{a}}$, and $ {\bf{A}}$, respectively. ${\left(  \cdot  \right)}^{ T}$, ${\left(  \cdot  \right)}^{ H}$,  ${\rm Tr}\left(  \cdot  \right)$  and ${\Re }(\cdot)$ denote  transpose, conjugate transpose, trace, and real dimension, respectively. ${\cal C}{\cal N}\left( {\mu,{\sigma ^2}} \right)$ denotes the distribution  of a \emph{circularly symmetric complex Gaussian} (CSCG) random variable with mean $\mu$ and variance $ {\sigma ^2}$. ${\mathbb C}^{x\times y}$ and ${\mathbb R}^{x\times y}$  denote the space of $x\times y$ complex/real matrices. ${\bm I}_K\in{\mathbb R}^{K\times K}$ is the identify matrix, ${\bm 1}_K=[1,\cdots,1]^T\in{\mathbb R}^{K\times 1}$, and $(a)^{+}\!=\!{\rm max}(0,a)$.

\section{System Model}\label{sec:System Model}
As shown in Fig.~\ref{fig:fig1}, we consider a programmable downlink MISO broadcast system which consists of one BS, one IRS, $K$ legitimate receivers, denoted as $B_1,\cdots,B_K$, and $N$ active eavesdroppers, denoted as $E_1,\cdots,E_N$. The BS and the IRS are equipped with $M$ antennas and $L$ reflecting elements, respectively, while the legitimate receivers and eavesdroppers are all equipped with a single antenna each.
The BS sends $K$ independent confidential data streams with one stream for each of the $K$ legitimate receivers over the same frequency band, simultaneously. At the same time, the unauthorized eavesdroppers are trying to eavesdrop any of the data streams, independently.

Consider the special case when the legitimate receivers and the eavesdroppers are in the same directions to the BS. In this case, the channel responses of the legitimate receivers will be highly correlated with those of the eavesdroppers. As aforementioned, it is intractable to guarantee the secret communications with the use of beamforming only at the transceivers.
Hence, we want to explore the use of the IRS to provide additional communication links so as to increase the SNR at the legitimate receivers while suppressing the SNR at the eavesdroppers. Hopefully, this will create an effect as if the confidential data streams can bypass the eavesdroppers and reach the legitimate receivers, and thus the secrecy rate will be improved.
In this paper, we are interested in obtaining the performance limit of such a system. Hence, similarly to \cite{Ruizhang2018Sufface} and \cite{huang2018large}, we assume that the CSI of all the channels are perfectly known at the BS\footnote{In practice, the optimization processing only requires the CSIs of the composite channels, i.e., ${\bm H}_k$ and ${\bm G}_n$, defined in \eqref{CC1} and \eqref{CC2}. Hence, we can first turn off the IRS and enable the users to send orthogonal pilot sequences to the BS for estimating the last rows of  ${\bm H}_k$ and ${\bm G}_n$. Next, we can turn on each reflecting element on the IRS successively in $L$ time slots and keep the other reflecting elements closed. Then, the users send orthogonal pilot sequences to the BS in the each time slot. Finally, the first $L$ rows in ${\bm H}_k$ and ${\bm G}_n$ can be estimated by subtracting the last rows of ${\bm H}_k$ and ${\bm G}_n$ from the estimated channel gains.}. In practical systems where such CSI cannot be obtained perfectly, the results derived in this paper can be considered as the performance upper bound.
Note that the optimization (in terms of beamformers and reflecting coefficients) of the system to be presented in the subsequent sections is done at the BS and that the optimized reflecting coefficients are transmitted to the IRS to reconfigure the corresponding reflecting elements accordingly.


\subsection{Channel  Model}
The baseband equivalent channel responses from the BS to the IRS, from the BS to $B_k$, from the BS to $E_n$, from the IRS to $B_k$, and from the IRS to $E_n$ are denoted by ${\bm F}\in {\mathbb C}^{L \times M}$, ${\bm h}_{d,k}^H\in {\mathbb C}^{1 \times M}$, ${\bm g}_{d,n}^H\in {\mathbb C}^{1 \times M}$, ${\bm h}_{r,k}^H\in {\mathbb C}^{1 \times L}$, and ${\bm g}_{r,n}^H\in {\mathbb C}^{1 \times L}$, respectively, with $1\le k\le K$ and $1\le n\le N$.
Specifically, without loss of generality, we adopt a Rician fading channel model, which consists of LoS and non-LoS (NLoS) components, i.e.,
\begin{align}
{\bm{h}} = \sqrt {\frac{{{\kappa _{\bm{h}}}}}{{{\kappa _{\bm{h}}} + 1}}} {{\bm{h}}^{\rm {LoS}}} + \sqrt {\frac{1}{{{\kappa _{\bm{h}}} + 1}}} {{\bm{h}}^{\rm{NLoS}}},\label{channel1}
\end{align}
with ${\bm{h}} \in {\bm {\cal H}}=\left\{ {\bm F} ,{{{\bm{h}}_{d,k}},{{\bm{h}}_{r,k}},{{\bm{g}}_{d,n}},{{\bm{g}}_{r,n}}} \right\}$, where ${\kappa _{\bm{h}}}$, ${{\bm{h}}^{\rm {LoS}}}$, and ${{\bm{h}}^{\rm {NLoS}}}$ are the Rician factor, LoS components, and NLoS components of channel $\bm h$, respectively.
The NLoS components ${{\bm{h}}^{\rm {NLoS}}}$ are i.i.d. complex Gaussian distributed with zero mean and unit variance. We define a vector ${{\bm{a}}_X}\left( \vartheta  \right) = \left[ {1,{e^{j\frac{{2\pi d}}{\lambda }\sin \vartheta }},\cdots,{e^{j\frac{{2\pi d}}{\lambda }\left( {X - 1} \right)\sin \vartheta }}} \right]^T$, where  $d$ is the antenna element separation, $\lambda$ is the carrier wavelength, $X$ is the dimension of the vector and $\vartheta$ is the angle, which can be interpreted as either {\emph{angle of departure}} (AoD) or {\emph{angle of arrival}} (AoA) depending on the context.  We set $d/\lambda = 1/2$ for simplicity. Hence, the LoS components in \eqref{channel1} can be modeled as
\begin{align}
\!\!\!\!{\bm{h}}_{d,k}^{\rm LoS} &\!=\! {{\bm{a}}_M}( {{\vartheta _{d,k}}} )\;{\rm and}\;{\bm{h}}_{r,k}^{\rm LoS}\! = \!{{\bm{a}}_L}( {{\vartheta _{r,k}}}), \;{\rm for}\; 1\le k\le K,\\
\!\!\!\!{\bm{g}}_{d,n}^{\rm LoS} &\!= \!{{\bm{a}}_M}( {{\tilde \vartheta _{d,n}}} )\;{\rm and}\;{\bm{h}}_{r,n}^{\rm LoS} \!= \!{{\bm{a}}_L}( {{\tilde \vartheta _{r,n}}}), \;{\rm for}\; 1\le n\le N,\\
\!\!\!\!{\bm{F}}^{\rm LoS} &\!= \!{{\bm{a}}_L}\left( {\vartheta ^{\rm AoA}} \right){\bm{a}}_M^H\left( {\vartheta ^{\rm AoD}} \right),\label{channel2}
\end{align}
where ${{\vartheta _{d,k}}}$, ${{\vartheta _{r,k}}}$, ${{\tilde \vartheta _{d,n}}}$, ${{\tilde \vartheta _{r,n}}}$  are the AoA or AoD of a signal from the BS to $B_k$, from the IRS to $B_k$, from the BS to $E_n$, and from the IRS to $E_n$, respectively. ${\vartheta^{\rm  AoD}}$ and ${\vartheta^{\rm  AoA}}$ are the AoD from the BS and the AoA to the IRS, respectively.

\subsection{Reflecting Coefficient Model}
The reflecting coefficient channel of the IRS \cite{Ruizhang2018Sufface} is given by
${\bm \Theta}={\text {  diag}}({\bm \theta})\in {\mathbb C}^{L\times L}$ with ${\bm \theta} =[\theta_1,\theta_2,\cdots,\theta_L]^T\in {\mathbb C}^{L\times 1}$ and $\theta_l \in {\bm {\Phi}}$ for $1\le l\le L$, where ${\text { diag}}(\cdot)$ denotes a diagonal matrix whose diagonal elements are given by the corresponding vector and $ {\bm {\Phi}}$ denotes the set of reflecting coefficients of the IRS. In this paper, we consider the following three different sets of reflecting coefficients, which lead to three different constraints for the reflecting coefficients.

\begin{itemize}
\item \emph{Continuous Reflecting Coefficients:} In this scenario, we further consider two detailed setups with the optimized or constant amplitude. Specifically, the reflecting coefficient set for the optimized amplitude with continuous phase-shift is denoted by
\begin{eqnarray}
{\bm {\Phi}}_1 = \left\{ {{\theta }_l\left| {{{\left| {{\theta }}_l \right|}^2} \le 1} \right.} \right\},
\end{eqnarray}
and the reflecting coefficient set for the constant amplitude with continuous phase-shift is denoted by
\begin{eqnarray}
{\bm {\Phi}}_2 = \left\{ {\theta_l\left|\theta_l={e^{j\varphi_l}},\varphi_l\in\left[0,2\pi\right)  \right.} \right\}.
\end{eqnarray}
\item \emph{Discrete Reflecting Coefficients:} In this scenario, the reflecting coefficient set has constant amplitude and discrete  phase-shift, which is given by
\begin{eqnarray}
{\bm {\Phi}}_3\! =\! \!\left\{ {\theta_l\left|\theta_l\!=\!{e^{j\varphi_l}},\!\varphi_l\!\in\!\!\left\{0,\textstyle{\frac{2\pi}{Q}},\cdots,\textstyle{\frac{2\pi(Q-1)}{Q}}\right\}  \right.} \right\},
\end{eqnarray}
where $Q$ is the number of reflecting coefficient values of the reflecting elements on the IRS.
\end{itemize}

Note that, it is costly in practice to achieve continuous reflecting coefficient on the reflecting elements due to the hardware limitation. Hence, applying the discrete reflecting coefficient on the reflecting elements, i.e., ${\bm \Phi}_3$, is more practical than applying the continuous reflecting coefficients, i.e., ${\bm \Phi}_1$ and ${\bm \Phi}_2$. But, it is also important to investigate the system performance with ${\bm \Phi}_1$ and ${\bm \Phi}_2$  since it serves as the upper bound to that with ${\bm \Phi}_3$.

\subsection{Signal Model}

Let ${s_k}$ be the confidential message dedicated to $B_k$.  It is assumed that all messages transmitted are CSCG, i.e., ${s_k}\sim {\cal C}{\cal N}\left( {0,1} \right)$ for $1\le k \le K$. Then, the signal transmitted from the BS can be expressed as
\begin{eqnarray}
{\bm{x}} = \sum\nolimits_{k = 1}^K {{{\bm{w}}_k}{s_{_k}}},
\end{eqnarray}
where ${{\bm{w}}_k}$ is the downlink beamforming vector for ${s_k}$. The received signals at $B_k$ and eavesdropped by $E_n$  can be  expressed as
\begin{eqnarray}
&y_k^{\rm{B}} = \left[ {{\bm{h}}_{r,k}^H{\bm \Theta} {\bm{F}} + {\bm{h}}_{d,k}^H} \right]\sum\limits_{i = 1}^K {{{\bm{w}}_i}{x_i}}  + u_k^{\rm{B}},1\le k\le K,\label{equ:SINRb}\\
&y_n^{\rm{E}} = \left[ {{\bm{g}}_{r,n}^H{\bm \Theta} {\bm{F}} + {\bm{g}}_{d,n}^H} \right]\sum\limits_{i = 1}^K {{{\bm{w}}_i}{x_i}}  + u_n^{\rm{E}},1\le n\le N,\label{equ:SINRee}
\label{equ:SINRe}
\end{eqnarray}
respectively, where $u_k^{\rm{B}}$ and $u_n^{\rm{E}}$ are  the received noises at $B_k$ and $E_n$, respectively.  It is assumed that all noises are Guassian distributed with zero mean, i.e., $u_k^{\rm{B}}\sim {\cal C}{\cal N}\left( {0,{\sigma_k ^2}} \right)$ and $u_k^{\rm{E}}\sim {\cal C}{\cal N}\left( {0,{\delta_n ^2}} \right)$, respectively.

According to (\ref{equ:SINRb}), the achievable transmission rate of the $k$-th confidential message received at $B_k$ can be written as
\begin{eqnarray}
R_k^{\rm{B}}  =\ln\left(1+ \frac{{{\left| {({{\bm{h}}_{r,k}^H{\bm \Theta} {\bm{F}} + {\bm{h}}_{d,k}^H}){\bm w}_k} \right|^2}}}{{\sum\nolimits_{i \ne k}^K {{\left|{({{\bm{h}}_{r,k}^H{\bm \Theta} {\bm{F}} + {\bm{h}}_{d,k}^H}){\bm w}_i } \right|^2} + {\sigma_k ^2}} }}\right).\label{eq7}
\end{eqnarray}

According to (\ref{equ:SINRee}), if $E_n$ attempts  to eavesdrop the $k$-th confidential message, the achievable wiretapped rate of the $k$-th message received at $E_n$ can be written as
\begin{eqnarray}
R_{k,n}^{\rm{E}}=\ln\left(1+ \frac{{{{\left|({{\bm{g}}_{r,n}^H{\bm \Theta} {\bm{F}} + {\bm{g}}_{d,n}^H}){\bm w}_k \right|}^2}}}{{\sum\nolimits_{i \ne k}^K {{{\left| ({{\bm{g}}_{r,n}^H{\bm \Theta} {\bm{F}} + {\bm{g}}_{d,n}^H}){\bm w}_i \right|}^2} + {\delta_n ^2}} }}\right).\label{eq8}
\end{eqnarray}

Since each eavesdropper can eavesdrop any of the $K$ confidential messages, the achievable secrecy rate (in nats/sec/Hz) for transmitting $s_k$ to $B_k$ and keeping it confidential from all the $N$ eavesdroppers should be the minimum-secrecy-rate among $B_k$ and $E_n$ for $1\le n \le N$, which is given by \cite{nasir2017secrecy}
\begin{eqnarray}
{C_k} = \mathop {\min }\limits_{\forall n} \left\{ R_k^{\rm{B}}- R_{k,n}^{\rm{E}}\right\}.
\end{eqnarray}

\section{Problem Statement }\label{secProblemStatement}
\subsection{Problem Formulation}
In this paper, we attempt to jointly optimize the beamfroming vector, i.e., ${\bm{W}} = [{{\bm{w}}_1},\cdots,{{\bm{w}}_k}]\in {\mathbb C}^{M\times K}$, and reflecting coefficients, i.e., $\bm \theta$, to maximize the minimum-secrecy-rate among all the legitimate receivers. Mathematically, the optimized problem can be generally formulated as

\begin{subequations}
\begin{align}
({\rm {\bf P}}1):\mathop {\max }\limits_{{\bm{W}},{\bm \theta} }\mathop {\min }\limits_{\forall k} \quad& {C_k}\nonumber \\
{\rm s.t.}\quad&\sum\nolimits_{k = 1}^K {{{\left\| {{{\bm{w}}_k}} \right\|}^2}}  \le P,\label{cons1}\\
&{\theta _l} \in {\bm{\Phi}}, 1\le l \le L\label{cons2},
\end{align}
\end{subequations}
where ${P}$  denotes the maximum transmit power at the BS and ${\bm{\Phi}}$ may be set as ${\bm{\Phi}}_1$, ${\bm{\Phi}}_2$, and ${\bm{\Phi}}_3$, respectively.
\subsection{Problem Transformation}
({\rm {\bf P}}1)  is hard to solve due to the non-concave objective function. In order to find the solution of ({\rm {\bf P}}1) efficiently, we will transform it into the following equivalent formulation.

To begin with, denoting ${\bm H}_k=\left[ {\begin{array}{*{20}{c}}{\text { diag}}({\bm{h}}_{r,k}^H){\bm F}\\{\bm{h}}_{d,k}^H\end{array}} \right]\in {\mathbb C}^{(L+1) \times M}$, and ${\bm G}_n=\left[ {\begin{array}{*{20}{c}}{\text { diag}}({\bm{g}}_{r,n}^H){\bm F}\\{\bm{g}}_{d,n}^H\end{array}} \right]\in {\mathbb C}^{(L+1) \times M} $,
we have
\begin{eqnarray}
&\left| {({{\bm{h}}_{r,k}^H{\bm \Theta} {\bm{F}} + {\bm{h}}_{d,k}^H}){\bm w}_k} \right|^2=\left| {\bm v}^H{\bm H}_k{\bm w}_k  \right|^2,\label{CC1}\\
&\left| {({{\bm{g}}_{r,n}^H{\bm \Theta} {\bm{F}} + {\bm{g}}_{d,n}^H}){\bm w}_k} \right|^2=\left| {\bm v}^H{\bm G}_n{\bm w}_k  \right|^2,\label{CC2}
\end{eqnarray}
where ${\bm v}=\left[v_1,v_2,\cdots,v_{L+1}\right]^T=\left[{\bm \theta} ; 1\right]\in {\mathbb C}^{1\times(L+1)} $.

Then, $R_k^{\rm{B}}$ in \eqref{eq7} and $R_{k,n}^{\rm{E}}$ in \eqref{eq8} can be rewritten as
\begin{align}
&R_k^{\rm{B}}=\ln\left(1+ \frac{{{\left|{\bm v}^H{\bm H}_k{\bm w}_k \right|^2}}}{b_{k}\left( {{\bm{W}},{\bm{v }}} \right)}\right) \buildrel \Delta \over =f_k^{\rm{B}}\left( {{\bm{W}},{\bm{v}}} \right), \\
&R_{k,n}^{\rm{E}}=\ln\left(1+ \frac{{{\left|{\bm v}^H{\bm G}_n{\bm w}_k \right|^2}}}{q_{k,n}\left( {{\bm{W}},{\bm{v }}} \right)}\right) \buildrel \Delta \over =f_{k,n}^{\rm{E}}\left( {{\bm{W}},{\bm{v}}} \right),
\end{align}
where ${b_{k}\left( {{\bm{W}},{\bm{v }}} \right)}={{\sum\nolimits_{i \ne k}^K {{\left|{\bm v}^H{\bm H}_k{\bm w}_i\right|^2} + {\sigma_k ^2}} }}$ and
${q_{k,n}\left( {{\bm{W}},{\bm{v }}} \right)}={{\sum\nolimits_{i \ne k}^K {{\left|{\bm v}^H{\bm G}_n{\bm w}_i\right|^2} + {\delta_n ^2}} }}$. Thus, it is straightforward to know that ({\rm {\bf P}}1) can be transformed into the following equivalent form:
\begin{subequations}
\begin{align}
\!\!\!\!({\rm {\bf P}}2):\mathop {\max }\limits_{{\bm{W}},{\bm v}}\quad &{\cal R}({\bm W},{\bm v})\buildrel \Delta \over ={\mathop {\min }\limits_{\forall k, \forall n} }\left\{f_{k}^{\rm{B}}\left( {{\bm{W}},{\bm{v}}} \right)-f_{k,n}^{\rm{E}}\left( {{\bm{W}},{\bm{v}}} \right)\right\}\nonumber \\
{\rm s.t.}
\quad&{v_l} \in {\bm{\Phi}}, 1\le l\le L, v_{L+1}=1,\label{consB}\\
&\eqref{cons1}.\nonumber
\end{align}
\end{subequations}

However, the transformed problem ({\rm {\bf P}}2) is still hard to solve  since ${\cal R}({\bm W},{\bm v})$ is not jointly concave with respect to ${\bm W}$ and $\bm v$, and even worse, they are coupled together.
In the next section, we will develop an iterative algorithm to solve ({\rm {\bf P}}2) efficiently.

\section{Minimum-Secrecy-Rate Maximization}\label{secProblemSolution}

In this section, we will propose two techniques to jointly solve the above challenging problem. Firstly, we apply the path-following algorithm to handle the non-concavity of the objective function. Then, we apply the alternating optimization  technique to deal with the coupled optimization variables. Finally, we analyze the convergence of the proposed  algorithm.

\subsection{Path-Following Algorithm Development}
In this part, we will develop path-following iterative algorithm to solve ({\rm {\bf P}}2) with the non-concave objective function, i.e., ${\cal R}({\bm W},{\bm v})$. In particular, the basic idea of the path-following is to follow a solution path of a family of the approximated problems of ({\rm {\bf P}}2). For example,  ${\cal R}({\bm W},{\bm v})$ is approximated by a concave lower bound function, which is obtained by applying linearly interpolating between the non-concave term $f_{k}^{\rm{B}}\left( {{\bm{W}},{\bm{v}}} \right)$ and the non-convex term $f_{k,n}^{\rm{E}}\left( {{\bm{W}},{\bm{v}}} \right)$, respectively. Specifically, the approximated problem has a local (global) optimal value and can be increased in each iteration, which finally leads to a local (global) optimal solution of ({\rm {\bf P}}2) \cite{zaslavskiy2009path}.

To begin with, let $( {{\bm{W}}^{(t)},{\bm{v}}^{(t)}} )$ denote the solution of ({\rm {\bf P}}2) in the $t$-th iteration. Then, in order to find the concave lower bound function of ${\cal R}({\bm W},{\bm v})$ to develop path-following algorithm, we can fist find the lower bound function of $f_{k}^{\rm{B}}( {{\bm{W}},{\bm{v}}} )$ and the  upper bound function of $f_{k,n}^{\rm{E}}( {{\bm{W}},{\bm{v}}} )$ at $( {{\bm{W}}^{(t)},{\bm{v}}^{(t)}})$. The details are given in the following lemma.

\begin{lemma}\label{seceqapproximationequal}
The  lower bound function of $f_{k}^{\rm{B}}\left( {{\bm{W}},{\bm{v}}} \right)$ and the  upper bound function of $f_{k,n}^{\rm{E}}\left( {{\bm{W}},{\bm{v}}} \right)$ at $\left( {{\bm{W}}^{(t)},{\bm{v}}^{(t)}} \right)$ in the $(t+1)$-th iteration of path-following algorithm are given by
\begin{align}
& f_k^{\rm{B}}( {{\bm{W}},{\bm{v}}} ) \ge
f_k^{\rm{B}}( {{{\bm{W}}^{^{( t )}}}\!\!\!,{{\bm{v}}^{^{( t )}}}} )\!+\!2\textstyle {\frac{{\Re \left\{ { {{{( {{{\bm{w}}_k^{( t )}}} )}^H}\!{{\bm{H}}_k^H}{\bm{v}}^{( t )}} \left( {{{\bm{v}}^H}{{\bm{H}}_k}{{\bm{w}}_k} } \right)} \right\}}}{{{b_k}( {{{\bm{W}}^{^{( t )}}},{{\bm{v}}^{^{( t )}}}} )}}}\nonumber\\
&\textstyle{- \frac{{ { {{\left| {{{( {{{\bm{v}}^{( t )}}} )}^H}{{\bm{H}}_k}{\bm{w}}_k^{( t )}} \right|}^2}} }}{{{b_k}( {{{\bm{W}}^{^{( t )}}},{{\bm{v}}^{^{( t )}}}} )( {{b_k}( {{{\bm{W}}^{^{( t )}}},{{\bm{v}}^{^{( t )}}}} ) + {{\left| {{{( {{{\bm{v}}^{( t )}}} )}^H}{{\bm{H}}_k}{\bm{w}}_k^{( t )}} \right|}^2}} )}}}\nonumber\\
&\times \textstyle{ \left( {{{\left| {{{\bm{v}}^H}{{\bm{H}}_k}{{\bm{w}}_k}} \right|}^2} \!+\! {b_k}( {{\bm{W}},{\bm{v}}} )} \right)- \frac{{{{\left| {{{\left( {{{\bm{v}}^{( t )}}} \right)}^H}{{\bm{H}}_k}{\bm{w}}_k^{( t )}} \right|}^2}}}{{{b_k}( {{{\bm{W}}^{^{( t )}}},{{\bm{v}}^{^{( t )}}}} )}}}\nonumber\\
&\buildrel \Delta \over = f_k^{\rm{B}}( {{\bm{W}},{\bm{v}};{{\bm{W}}^{^{( t )}}},{{\bm{v}}^{^{( t )}}}} ),\label{path1}\\
&f_{k,n}^{\rm{E}}( {{\bm{W}},{\bm{\theta }}} )
\le\textstyle{ f_{k,n}^{\rm{E}}( {{{\bm{W}}^{^{( t )}}},{{\bm{v}}^{^{( t )}}}} )\!  + \! ( {1\!  + \! \frac{{{{\left| {{{( {{{\bm{v}}^{( t )}}} )}^H}{{\bm{G}}_n}{\bm{w}}_k^{( t )}} \right|}^2}}}{{{q_{k,n}}( {{{\bm{W}}^{^{( t )}}},{{\bm{v}}^{^{( t )}}}} )}}})^{ - 1}}\nonumber\\
&\times\textstyle{(  \frac{{{{\left| {{{\bm{v}}^H}{{\bm{G}}_n}{{\bm{w}}_k}} \right|}^2}}}{{{q_{k,n}}( {{\bm{W}},{\bm{v}}} )}} -   \frac{{{{\left| {{{( {{{\bm{v}}^{( t )}}} )}^H}{{\bm{G}}_n}{\bm{w}}_k^{( t )}} \right|}^2}}}{{{q_{k,n}}( {{{\bm{W}}^{^{( t )}}},{{\bm{v}}^{^{( t )}}}} )}}  )}\nonumber\\
&\textstyle{\le f_{k,n}^{\rm{E}}( {{{\bm{W}}^{^{( t )}}},{{\bm{v}}^{^{( t )}}}} ) + ( {1 + \frac{{{{\left| {{{( {{{\bm{v}}^{( t )}}} )}^H}{{\bm{G}}_n}{\bm{w}}_k^{( t )}} \right|}^2}}}{{{q_{k,n}}( {{{\bm{W}}^{^{( t )}}},{{\bm{v}}^{^{( t )}}}} )}}})^{ - 1}}\nonumber\\
&\times\textstyle{(  \frac{{{{\left| {{{\bm{v}}^H}{{\bm{G}}_n}{{\bm{w}}_k}} \right|}^2}}}{{{{q_{k,n}}( {{\bm{W}},{\bm{v}};{{\bm{W}}^{^{( t )}}},{{\bm{v}}^{^{( t )}}}} )}}} -   \frac{{{{\left| {{{( {{{\bm{v}}^{( t )}}} )}^H}{{\bm{G}}_n}{\bm{w}}_k^{( t )}} \right|}^2}}}{{{q_{k,n}}( {{{\bm{W}}^{^{( t )}}},{{\bm{v}}^{^{( t )}}}} )}}  )}\nonumber\\
&\buildrel \Delta \over = f_{k,n}^{\rm{E}}( {{\bm{W}},{\bm{v}};{{\bm{W}}^{^{( t )}}},{{\bm{v}}^{^{( t )}}}} ),\label{path2}
\end{align}
where 
\begin{align}
&{q_{k,n}}( {{\bm{W}},{\bm{v}};{{\bm{W}}^{^{( t )}}},{{\bm{v}}^{^{( t )}}}} )= {\delta_n ^2}+\nonumber\\
&\textstyle{\sum\nolimits_{i \ne k}^K {\!\!\Re\! \{\!{( {{\bm w}_i^{(t)}} )^H}{\bm G}_n^H{{\bm v}^{(t)}}\!
( {2{{\bm{v}}^H}\! {{\bm{G}}_n}{{\bm{w}}_i} \!- \!{{( {{{\bm{v }}^H}} )}^{( \!t \!)}}\! {{\bm{G}}_n}{\bm{w}}_i^{^{(\! t\! )}}} ) \}  }},\label{path3}
\end{align}
\end{lemma}
\begin{IEEEproof}
 Please refer to Appendix \ref{throem41}.
 \end{IEEEproof}

Then, from \eqref{path1} and \eqref{path2}, we know the lower bound of ${\cal R}({\bm W},{\bm v})$ is given by
\begin{align}
& {\cal R}({\bm W},{\bm v})={\mathop {\min }\limits_{\forall k, \forall n} }\left\{f_{k}^{\rm{B}}\left( {{\bm{W}},{\bm{v}}} \right)-f_{k,n}^{\rm{E}}\left( {{\bm{W}},{\bm{v}}} \right)\right\}\nonumber\\
&\ge{\mathop {\min }\limits_{\forall k, \forall n} }\left\{f_k^{\rm{B}}( {{\bm{W}},{\bm{v}};{{\bm{W}}^{^{( t )}}},{{\bm{v}}^{^{( t )}}}} ) - f_{k,n}^{\rm{E}}( {{\bm{W}},{\bm{v}};{{\bm{W}}^{^{( t )}}},{{\bm{v}}^{^{( t )}}}})\right\}  \nonumber\\
&\buildrel \Delta \over ={\cal R}^{\rm lb}( {{\bm{W}},{\bm{v}};{{\bm{W}}^{^{( t )}}},{{\bm{v}}^{^{( t )}}}}).\label{eq19}
\end{align}
Note that, according to \eqref{path1} and \eqref{path2}, the equality in \eqref{eq19} holds when ${\bm W}={\bm W}^{(t)}$ and ${\bm v}={\bm v}^{(t)}$.

Thus, a family of the approximated problems of ({\rm {\bf P}}2) is given as follows:
\begin{subequations}
\begin{align}
\!\!({\rm {\bf P}}2\!-\!t):\mathop {\max }\limits_{{\bm{W}},{\bm v}}\quad \!\!&{\cal R}^{\rm lb}( {{\bm{W}},{\bm{v}};{{\bm{W}}^{^{( t )}}},{{\bm{v}}^{^{( t )}}}}) \nonumber \\
{\rm s.t.}\quad\!\!
&\eqref{cons1}  \; {\rm and}\; \eqref{consB}.\nonumber
\end{align}
\end{subequations}
However, $({\rm {\bf P}}2-t)$ is still a non-convex problem due to the following reasons:
\begin{itemize}
\item First, $\bm{W}$ and $\bm{v} $ are coupled in the terms of ${{{\bm{v}}^H}{{\bm{H}}_k}{{\bm{w}}_k}}$ and ${{{\bm{v}}^H}{{\bm{G}}_n}{{\bm{w}}_k}}$, which makes the objective function ${\cal R}^{\rm lb}( {{\bm{W}},{\bm{v}};{{\bm{W}}^{^{( t )}}},{{\bm{v}}^{^{( t )}}}})$ not jointly concave with respect to $({\bm W},{\bm v})$.
\item Second, it is straightforward to know that \eqref{consB} with ${\bm{\Phi}}={\bm{\Phi}}_1$  is a convex set but  a non-convex set with ${\bm{\Phi}}={\bm{\Phi}}_2$ and ${\bm{\Phi}}={\bm{\Phi}}_3$.
\end{itemize}

In subsection \ref{subsecB}, we will first develop alternating optimization method to deal with the coupled optimization variables in $({\rm {\bf P}}2\!-\!t)$ with ${\bm{\Phi}}={\bm{\Phi}}_1$, and then we will extend it to the scenarios with ${\bm{\Phi}}={\bm{\Phi}}_2$ and ${\bm{\Phi}}={\bm{\Phi}}_3$, respectively.

\subsection{Alternating Optimization with Continuous and Discrete Reflecting Coefficients}\label{subsecB}
\subsubsection{The Solution of ({\rm {\bf P}2}) with ${\bm{\Phi}}={\bm{\Phi}}_1$}
In this part, we develop the alternating optimization to solve $({\rm {\bf P}}2-t)$ when ${\bm{\Phi}}={\bm{\Phi}}_1$ in constraint \eqref{consB}, which leads constraint \eqref{consB} to be a convex set. Hence, the non-convexity of $({\rm {\bf P}}2-t)$ only stems from the coupled optimization variables.


In fact, although the objective function ${\cal R}^{\rm lb}( {{\bm{W}},{\bm{v}};{{\bm{W}}^{^{( t )}}},{{\bm{v}}^{^{( t )}}}})$ is non-concave due to the coupled $\bm{W}$ and $\bm{v} $,  $f_k^{\rm{B}}( {{\bm{W}},{\bm{v}};{{\bm{W}}^{^{( t )}}},{{\bm{v}}^{^{( t )}}}} )$ in \eqref{path1} is biconcave in $\bm W$ and $\bm v$, i.e., $f_k^{\rm{B}}( {{\bm{W}},{\bm{v}};{{\bm{W}}^{^{( t )}}},{{\bm{v}}^{^{( t )}}}} )$  is  concave both in $\bm W$ with fixed $\bm v$  and in $\bm v$ with fixed $\bm W$.
Similarly, for the domain ${q_{k,n}}( {{\bm{W}},{\bm{v}};{{\bm{W}}^{^{( t )}}},{{\bm{v}}^{^{( t )}}}} )\ge0$, the function $ \frac{{{{\left| {{{\bm{v}}^H}{{\bm{G}}_n}{{\bm{w}}_k}} \right|}^2}}}{{{{q_{k,n}}( {{\bm{W}},{\bm{v}};{{\bm{W}}^{^{( t )}}},{{\bm{v}}^{^{( t )}}}} )}}}$ in \eqref{path2} is a biconvex function with respect to $\bm W$ and $\bm v$, which leads to a biconvex function $f_{k,n}^{\rm E}( {{\bm{W}},{\bm{v}};{{\bm{W}}^{^{( t )}}},{{\bm{v}}^{^{( t )}}}})$ with respect to $\bm W$ and $\bm v$. Hence, ${\cal R}^{\rm lb}( {{\bm{W}},{\bm{v}};{{\bm{W}}^{^{( t )}}},{{\bm{v}}^{^{( t )}}}})$ is a biconcave function in $\bm W$ and $\bm v$.


Therefore, we know $({\rm {\bf P}}2-t)$ with ${\bm{\Phi}}={\bm{\Phi}}_1$ has convex constraints and concave objective function in  $\bm W$ with fixed $\bm v$ and in $\bm v$ with fixed $\bm W$. Hence, we can apply the alternating optimization method to solve  $({\rm {\bf P}}2-t)$ in an alternating manner efficiently. Specifically, the alternating algorithm decouples $({\rm {\bf P}}2-t)$ into the following two subproblems for the optimization of $\bm{W}$ and $\bm{v}$, respectively,
\begin{align}
({\rm {\bf P}}3\!-\!{\rm A}):\mathop {\max }\limits_{{\bm{W}} }\quad &{\cal R}^{\rm lb}( {{\bm{W}},{\bm{v}};{{\bm{W}}^{^{( t )}}},{{\bm{v}}^{^{( t )}}}}) \nonumber \\
{\rm s.t.}\quad&\eqref{cons1},\nonumber 
\end{align}
and
\begin{align}
\!\!({\rm {\bf P}}3\!-\!{\rm B}):\mathop {\max }\limits_{{\bm v} }\quad &{\cal R}^{\rm lb}( {{\bm{W}},{\bm{v}};{{\bm{W}}^{^{( t )}}},{{\bm{v}}^{^{( t )}}}}) \nonumber \\
{\rm s.t.}\quad&  \eqref{consB}\; {\rm with}\;{\bm{\Phi}}={\bm{\Phi}}_1.\nonumber
\end{align}
Note that $({\rm {\bf P}}3-{\rm A})$ is an  optimization subproblem for solving $\bm{W}$ with a given $\bm v$ and $({\rm {\bf P}}3-{\rm B})$ is an optimization subproblem for solving $\bm v$ with a given $\bm{W}$.

As aforementioned, we know both $({\rm {\bf P}}3-{\rm A})$ and $({\rm {\bf P}}3-{\rm B})$ are convex optimization problems, which can be solved optimally and efficiently by using CVX \cite{grant2015cvx1}. Thus, problem $({\rm {\bf P}}2)$ with ${\bm{\Phi}}={\bm{\Phi}}_1$ can be solved efficiently by alternately solving $({\rm {\bf P}}3-{\rm A})$ and $({\rm {\bf P}}3-{\rm B})$ in an iterative manner of path-following algorithm. In particular, the algorithm steps of the alternating optimization based path-following algorithm are summarized in Algorithm \ref{algorithm1}.
\begin{algorithm}
\caption{Alternating optimization based path-following algorithm}\label{algorithm1}
\small
{{
\begin{algorithmic}[1]
 \STATE Initialize ${\bm W}^{(0)}$, ${\bm v}^{(0)}$ and $t=0$.
\REPEAT
  \STATE $t \leftarrow t+1$,
 \STATE Set ${\bm v}={\bm v}^{(t-1)}$ and calculate ${\bm W}^{(t)}$ by solving the convex optimization problem $({\rm {\bf P}}3-{\rm A})$,
 \STATE Set ${\bm W}={\bm W}^{(t)}$ and calculate ${\bm v}^{(t)}$ by solving the convex optimization problem $({\rm {\bf P}}3-{\rm B})$,
 \UNTIL $\Gamma = \frac{\left({\cal R}( {{{\bm{W}}^{(t)}},{{\bm{v}}^{(t)}}} )-{\cal R}( {{{\bm{W}}^{(t-1)}},{{\bm{v}}^{(t-1)}}} )\right)}{{\cal R}( {{{\bm{W}}^{(t)}},{{\bm{v}}^{(t)}}} )}$ converges.
\end{algorithmic}
}}
\end{algorithm}
\subsubsection{The Solutions of  ({\rm {\bf P}2}) with  ${\bm{\Phi}}={\bm{\Phi}}_2$}\label{subsecP2}
In this part, we extend the above alternating optimization to solve $({\rm {\bf P}}2-t)$ when
${\bm{\Phi}}={\bm{\Phi}}_2$ in constraint \eqref{consB}, which leads constraint \eqref{consB} to be a non-convex set.
To handle this non-convex constraint, we propose the following two methods:

\begin{itemize}
\item In the first method, we introduce a positive constant relaxation factor $\lambda$ to reformulate $({\rm {\bf P}}2-t)$ with ${\bm{\Phi}}={\bm{\Phi}}_2$ as the following optimization problem,
\begin{subequations}
\begin{align}
\!\!({\rm {\bf P}}4\!-\!t):\mathop {\max }\limits_{{\bm{W}},{\bm v} }\quad\!\! &{\cal R}^{\rm lb}( {{\bm{W}},{\bm{v}};{{\bm{W}}^{^{( t )}}},{{\bm{v}}^{^{( t )}}}}) +\lambda{\sum\limits_{l = 1}^{L+ 1} {{{\left| {{v_l}} \right|}^2}}}\nonumber \\
{\rm s.t.}\quad& \eqref{cons1} \; {\rm and}\; \eqref{consB} \;{\rm with}\; {\bm{\Phi}}={\bm{\Phi}}_1.\nonumber
\end{align}
\end{subequations}
Note that  the added nonnegative quadratic term $\textstyle{\lambda{\sum\limits_{l = 1}^{L + 1} {{{\left| {{v_l}} \right|}^2}}}}$ attempts to force the inequality holds for ${v _l}$, i.e., ${\left| v \right|_l} = 1$.

$\quad$However, the objective of $({\rm {\bf P}}4-t)$ is to maximize the summation of concave and convex functions, which belongs to a non-convex  problem. To further deal with this challenge, we use the first-order Taylor series expansion to approximates the convex function as an affine function \cite{chen2018resource}.
Then, we iteratively solve the approximated convex optimization problem until the convergence is met. Specifically, the approximated  problem is
\begin{subequations}
\begin{align}
\!\!({\rm {\bf P}}4-{\rm A}):\mathop {\max }\limits_{{\bm{W}},{\bm v} }\quad\!\! &{\cal R}^{\rm lb}( {{\bm{W}},{\bm{v}};{{\bm{W}}^{^{( t )}}},{{\bm{v}}^{^{( t )}}}}) \nonumber\\
&+\lambda\sum\nolimits_{l= 1}^{L + 1} {\Re \left\{ {{{( {v_l^{(t)}} )}^H}( {2{v_l} - {{ {{v_l^{(t)}}} }}} )} \right\}} \nonumber \\
{\rm s.t.}\quad& \eqref{cons1}, \; {\rm and}\; \eqref{consB} \;{\rm with}\; {\bm{\Phi}}={\bm{\Phi}}_1.\nonumber
\end{align}
\end{subequations}

$\quad$Finally, the only non-convex term in $({\rm {\bf P}}4-{\rm A})$ stems from the coupled $\bm{W}$ and $\bm{v}$ in  the objective function ${\cal R}^{\rm lb}( {{\bm{W}},{\bm{v}};{{\bm{W}}^{^{( t )}}},{{\bm{v}}^{^{( t )}}}})$, which can be solved efficiently by applying the same alternating optimization method, i.e., Algorithm \ref{algorithm1}.

$\quad$However, this method has the main drawbacks that we use the approximation in $({\rm {\bf P}}4\!-\!t)$ and there is no beforehand choice for the relaxation factor $\lambda$ to speed up the convergence \cite{nasir2017secrecy}. Hence, we further propose the direct projection method in the next part.
\item In the second method, we can apply the projection method to project the solution of $({\rm {\bf P}}2)$ with ${\bm{\Phi}}={\bm{\Phi}}_1$ into ${\bm{\Phi}}={\bm{\Phi}}_2$ directly. Specifically, denote the solutions  of $({\rm {\bf P}}2)$ with ${\bm{\Phi}}={\bm{\Phi}}_1$ and $({\rm {\bf P}}2)$ with ${\bm{\Phi}}={\bm{\Phi}}_2$  as $({\bm W}^{\dag},{\bm v}^{\dag})$ and $({\bm W}^{\ddag},{\bm v}^{\ddag})$, respectively. Thus, the $({\bm W}^{\ddag},{\bm v}^{\ddag})$ can be obtained by solving the following projection problem:

\begin{align}
\!\!({\rm {\bf P}}4-{\rm B}):{\bm v}^{\ddag}=\mathop {\arg\min }\limits_{\bm{v}}\quad& \left\| {{\bm{v}} - {\bm{v}}^{\dag}} \right\|^2 \nonumber\\
{\rm s.t.}\quad&\eqref{consB} \;{\rm with}\; {\bm{\Phi}}={\bm{\Phi}}_2.\nonumber
\end{align}
From \cite{nadeem2019largearxiv}, the optimal solution to ({\rm {\bf P}}4-{\rm B}) is given by
\begin{align}
{\bm v}^{\ddag}=\exp(j\arg({\bm v}^{\dag})),\label{projection1}
\end{align}
and ${\bm W}^{\ddag}={\bm W}^{\dag}$. Note that, to further improve the performance of this method, we can rerun the following adjusted Algorithm \ref{algorithm1} to iteratively update the obtained $({\bm W}^{\ddag},{\bm v}^{\ddag})$:
\begin{itemize}
 \item Initialize ${\bm W}^{0}={\bm W}^{\ddag}$, ${\bm v}^{0}={\bm v}^{\ddag}$ and $t=0$. Then, perform the following steps iteratively until the objective function converges.
 \item $t \leftarrow t+1$, set ${\bm v}={\bm v}^{(t-1)}$ and calculate ${\bm W}^{(t)}$ by solving the convex optimization problem $({\rm {\bf P}}3-{\rm A})$,
 \item Set ${\bm W}={\bm W}^{(t)}$, project the solution of $({\rm {\bf P}}3-{\rm B})$ into ${\bm{\Phi}}_2$  by \eqref{projection1}, and denote it as ${\bm{\tilde v}}$,
 \item  Update ${\bm v}^{(t)}$ using the following rule:
 \begin{align}
\!\!\!\!\!\!\!\!\!\!{{\bm{v}}^{( t )}}  = \left\{ {\begin{array}{*{20}{l}}
\!\!{{\bm{\tilde v}}}, \;{\rm if}\;  {{\cal R}( {{{\bm{W}}^{( t )}},{\bm{\tilde v}}} )\ge{\cal R}( {{{\bm{W}}^{( t )}},{{\bm{v}}^{( {t - 1} )}}} )},\\
\!\!{{{\bm{v}}^{( {t - 1} )}}},\; {\rm otherwise}.
\end{array}} \right.
\end{align}
\end{itemize}
%
%
\end{itemize}

\subsubsection{The Solution of  ({\rm {\bf P}2})  with  ${\bm{\Phi}}={\bm{\Phi}}_3$}
In this part, we develop algorithms to solve $({\rm {\bf P}}2)$ when ${\bm{\Phi}}={\bm{\Phi}}_3$ in constraint \eqref{consB}, which leads the optimized problem  belongs to a class of combinatorial optimization problem, which is an NP-hard problem in general. Thus, it will cause intractable complexity to obtain the optimal solution. Hence, we will use the similar  heuristic projection  method in the above to solve this problem efficiently.


To begin with, we denote the solution of ${\bm{\Phi}}={\bm{\Phi}}_3$ as $({\bm W}^{\S}, {\bm v}^{\S})$. Then, we can directly project  ${\bm v}^{\dag}$, the solution of  ({\rm {\bf P}2})  with  ${\bm{\Phi}}={\bm{\Phi}}_1$, into   ${\bm{\Phi}}_3$ to obtain $({\bm W}^{\S}, {\bm v}^{\S})$, i.e.,
\begin{align}
\!\!\!v_l^\S \!= \!\!\left\{\!\!\!\! {\begin{array}{*{20}{l}}
{e^{j{\varphi _{\hat q}}}},\;{\rm where}\;\hat q \!=\!\mathop { \arg \min }\limits_{1 \le q \le Q} \left| {v_l^\dag \! - \!{e^{j{\varphi _q}}}} \right|, 1\le\! l\!\le L,\\
1,l=L+1,
\end{array}} \right.
\end{align}
where $v_l^\S $ and $v_l^\dag$ are the $l$-th element of ${\bm v}^\S$ and ${\bm v}^\dag$, respectively. ${\bm W}^{\S}={\bm W}^{\dag}$. Note that the rest steps to update $({\bm W}^{\S}, {\bm v}^{\S})$ are similar as the second method in subsection \ref{subsecP2}, which is omitted here for brevity.

\subsection{Convergence Analysis}
In this part, we analyze the convergence of the proposed alternating optimization based path-following algorithm, i.e., Algorithm \ref{algorithm1}, which is given in the following theorem.
\begin{theorem}\label{theorem1}
The value of the objective function increases in each iteration of Algorithm \ref{algorithm1}, i.e., ${\cal R}({\bm W}^{(t)},{\bm v}^{(t)})\le  {\cal R}({\bm W}^{(t+1)},{\bm v}^{(t+1)})$, which guarantees to converge to a local (global) optimum.
\end{theorem}
\begin{IEEEproof}
To begin with, we have
\begin{align}
{\cal R}( {{{\bm{W}}^{(t)}},{{\bm{v}}^{(t)}}} ) &\mathop  = \limits_{({\rm{a}})} {{\cal R}^{{\rm{lb}}}}( {{{\bm{W}}^{(t)}},{{\bm{v}}^{(t)}};{{\bm{W}}^{(t)}},{{\bm{v}}^{(t)}}} )\nonumber\\
&\le \left\{ {\mathop {\max }\limits_{\bm{W}} {{\cal R}^{{\rm{lb}}}}( {{\bm{W}},{{\bm{v}}^{(t)}};{{\bm{W}}^{(t)}},{{\bm{v}}^{(t)}}} )} \right\}\nonumber\\
&\mathop  = \limits_{({\rm{b}})} {{\cal R}^{{\rm{lb}}}}( {{{\bm{W}}^{(t + 1)}},{{\bm{v}}^{(t)}};{{\bm{W}}^{(t)}},{{\bm{v}}^{(t)}}} )\nonumber\\
&\le \left\{ {\mathop {\max }\limits_{\bm{v}} {{\cal R}^{{\rm{lb}}}}( {{{\bm{W}}^{(t + 1)}},{\bm{v}};{{\bm{W}}^{(t)}},{{\bm{v}}^{(t)}}} )} \right\}\nonumber\\
& \mathop = \limits_{({\rm{c}})} {{\cal R}^{{\rm{lb}}}}( {{{\bm{W}}^{(t + 1)}},{{\bm{v}}^{(t + 1)}};{{\bm{W}}^{(t)}},{{\bm{v}}^{(t)}}} )\nonumber\\
&\mathop \le \limits_{({\rm{d}})}{\cal R}( {{{\bm{W}}^{(t + 1)}},{{\bm{v}}^{(t + 1)}}} ),
\end{align}
where ({\rm{a}}) is because the equality in \eqref{eq19} holds when ${\bm W}={\bm W}^{(t)}$ and ${\bm v}={\bm v}^{(t)}$, ({\rm{b}}) and ({\rm{c}}) are because ${\bm W}^{(t + 1)}$ and ${\bm v}^{(t+1)}$ are the optimal solutions of the convex optimization problems of $(\rm{ {\bf P}}3-A)$ and $(\rm{ {\bf P}}3-B)$, respectively, and ({\rm{d}}) is because ${{\cal R}^{{\rm{lb}}}}( {{{\bm{W}}^{(t + 1)}},{{\bm{v}}^{(t + 1)}};{{\bm{W}}^{(t)}},{{\bm{v}}^{(t)}}} )$ is the lower bound of the function ${\cal R}( {{{\bm{W}}},{{\bm{v}}}} )$ in \eqref{eq19}.

Furthermore, due to \eqref{cons1} and \eqref{consB}, we know ${\bm W}^{(t)}$ and ${\bm v}^{(t)}$ are both bounded. According to Cauchy's theorem \cite{nasir2017secrecy}, we know the sequence of $({\bm W}^{(t)},{\bm v}^{(t)})$ will converge to $({\bm W}^*,{\bm v}^*)$ as $t\to \infty$, i.e.,
\begin{align}
0 &=\mathop {\lim }\limits_{t \to \infty } \left\{ {{\cal R}( {{{\bm{W}}^{(t)}},{{\bm{v}}^{(t)}}} ) - {\cal R}( {{{\bm{W}}^*},{{\bm{v}}^*}} )} \right\}\nonumber\\
 &\le \mathop {\lim }\limits_{t \to \infty } \left\{ {{\cal R}( {{{\bm{W}}^{(t + 1)}},{{\bm{v}}^{(t + 1)}}} ) - {\cal R}( {{{\bm{W}}^*},{{\bm{v}}^*}} )} \right\} = 0.
 \end{align}

Hence, we have proved the ${\cal R}({\bm W}^{(t)},{\bm v}^{(t)})\le  {\cal R}({\bm W}^{(t+1)},{\bm v}^{(t+1)})$, which can guarantee to converge to a local (global) optimum.
\end{IEEEproof}
\begin{theorem}\label{theorem2}
The corresponding solution $({\bm W}^{*},{\bm v}^{*})$ will converge to a {\emph{Karush-Kuhn-Tucker}} (KKT)  point finally.
\end{theorem}
\begin{IEEEproof}
Please refer to Appendix \ref{appendix2}
\end{IEEEproof}

However, although the convergence of the proposed algorithm is guaranteed, it requires solving the convex optimization problems $
({\rm {\bf P}}3-{\rm A})$ and $({\rm {\bf P}}3-{\rm B})$ whose complexity are in the order of ${\cal O}((KN+1)K^2M^2)$ and ${\cal O}((KN+L+1)(L+1)^2)$ \cite{nesterov1994interior}, respectively. Hence, we will develop the methods to reduce the computational complexity in the next subsection. Note that, since the solutions with ${\bm \Phi}_1$ can be projected into ${\bm \Phi}_2$ and ${\bm \Phi}_3$ directly, thus we only consider ${\bm \Phi}={\bm \Phi}_1$ when developing the low-complexity algorithms.

\section{Suboptimal Algorithms with Low-Complexity}\label{secLowcomplexity}
In this section, two suboptimal algorithms are developed to further reduce the complexity. Firstly, we develop an alternating optimization algorithm for the case with one legitimate user and one eavesdropper, where the closed-form solutions are provided in each iteration. Then, we develop an non-iterative suboptimal algorithm based on ZF beamforming for the case with multiple legitimate users and eavesdroppers.

\subsection{Alternating Optimization for ({\rm {\bf P}}2) with $K=1$ and $N=1$}
In this section, we develop a low-complexity algorithm to solve ({\rm {\bf P}}2) for the case with $K=1$ and $N=1$. Although there is no-interference in the objective function, the problem is still non-convex and hard to solve due to the coupled $\bm W$ and $\bm v$. Thus,  we also need to apply alternating optimization to decouple $\bm W$ and $\bm v$. Fortunately, we can obtain the closed-form solutions in each iteration, which leads it to be a low-complexity algorithm.

 Specifically, ({\rm {\bf P}}2) with $K=1$ and $N=1$ can be solved by alternately solving the following two subproblems:
\begin{align}
({\rm {\bf P}}5\!-\!{\rm A}):\mathop {\max }\limits_{{{\bm{w}}_1}} \quad&\ln \left({1 + {{| {{\tilde {\bm h}_1}^H{{\bm{w}}_1}} |}^2}} \right) - \ln \left( {1 + {{| {{\bm{\tilde g}}_1^H{{\bm{w}}_1}} |}^2}} \right)\nonumber\\
{\rm s.t.}\quad& {\left\| {\bm{w}}_1 \right\|^2} \le P,
\end{align}
which is an optimization problem of ${\bm w}_1$ for a given $\bm v$ with ${\tilde {\bm h}_1}^H={\bm v}^H{\bm H}_1$ and ${\tilde {\bm g}_1}^H={\bm v}^H{\bm G}_1$, and
\begin{align}
({\rm {\bf P}}5\!-\!{\rm B}):\mathop {\max }\limits_{{{\bm{v}}}}\quad&\ln \left( {1 + {{\left| {{\bm{v}}}^H{\bar {\bm h}_1} \right|}^2}} \right) - \ln \left( {1 + {{\left| {{{\bm{v}}}^H{\bm{\bar g}}_1} \right|}^2}} \right)\nonumber\\
{\rm s.t.}\quad&{\left| {{v_l}} \right|^2} \le 1, 1\le l \le L+1, \label{eqcons1}
\end{align}
which is an optimization problem of ${\bm v}$ for a given ${\bm w}_1$ with ${\bar {\bm h}_1}={\bm H}_1{\bm w}_1$ and ${\bm {\bar g}_1}= {\bm G}_1{\bm w}_1$.
Note that in constraint \eqref{eqcons1}, we have relaxed constraint \eqref{consB} due to $v_{L+1}$. To make the solution of $\bm v$ in the above problem  ({\rm {\bf P}}5-{\rm B}) to satisfy \eqref{consB}, we need to set $v_{L+1}^*$ as $1$ and $v_{l}^*=v_{l}/\exp(j\arg({ v_{L+1}}))$ for $1\le l\le L$ after the convergence. Besides, constraints \eqref{eqcons1} can be regarded as the per-antenna power constraints with the maximum power of one \cite{ZhangLan2012Gaussian}.

In the following two parts, we provide the solutions to ({\rm {\bf P}}5-{\rm A}) and ({\rm {\bf P}}5-{\rm B}), respectively.
\subsubsection{The Optimal Solution to ({\rm {\bf P}5}-{\rm A})}
This problem is the downlink MISO beamforming problem for basic wiretap channel, which has been studied in \cite{shafiee2007achievable}. In particular, the optimal solution to ({\rm {\bf P}}5-{\rm A}) is given by
\begin{align}
{\bm{w}}_1^* = \sqrt{P}\frac{{{{\left( {{\bm{I}} + P{{{\bm{\tilde g}}}_1}{\bm{\tilde g}}_1^H} \right)}^{ - \frac{1}{2}}}{\bm{q}}}}{{\left\| {{{\left( {{\bm{I}} + P{{{\bm{\tilde g}}}_1}{\bm{\tilde g}}_1^H} \right)}^{ - \frac{1}{2}}}{\bm{q}}} \right\|}},\label{eqopw1}
\end{align}
where ${{\bm{q}}}$ is the eigenvector of matrix $\bm Z$ corresponding to the largest eigenvalue, and
\begin{align}
{\bm{Z}}={( {{\bm{I}} + P{{{\bm{\tilde g}}}_1}{\bm{\tilde g}}_1^H} )^{ - \frac{1}{2}}}( {{\bm{I}} + P{{{\bm{\tilde h}}}_1}{\bm{\tilde h}}_1^H} )( {{\bm{I}} + P{{{\bm{\tilde g}}}_1}{\bm{\tilde g}}_1^H} )^{ - \frac{1}{2}}.
\end{align}

\subsubsection{The Solutions to ({\rm {\bf P}5}-{\rm B}) }
In this part, we develop the efficient algorithm to solve ({\rm {\bf P}}5-{\rm B}) in an iterative manner\footnote{Note that although ({\rm {\bf P}}5-B) can be solved by transforming it into a convex semidefinite programming (SDP) by applying semidefinite relaxation (SDR) technique, but solving the SDP problems requires very high complexity, which is in the order
of ${\cal O}((N+1)^{4.5})$ \cite{Chen2016}.}, and we provide the closed-form solutions in each iteration. To do so, we first introduce the following lemma
\begin{lemma}\label{theorem3}
Let $x$ be a positive real number, and define $f(y)=-xy+\ln y+1$, then we have
$-\ln  x  = \mathop { \max }\nolimits_{y > 0} f\left( y \right)$, and the optimal corresponding solution in the right hand side of this equation is $y^*=1/x$.
\end{lemma}
\begin{IEEEproof}
This proof is similar in \cite{jose2011robust}, which is omitted here for brevity.
\end{IEEEproof}

According to lemma \ref{theorem3},   ({\rm {\bf P}}5-{\rm B}) can be equivalently rewritten as
\begin{align}
({\rm {\bf P}}6):
\mathop {\max }\limits_{{\bm{v}},y} \quad & \ln(1 \!+\! \left| {{{\bm{v}}^H}{\bar{\bm{h}}_1}} \right|^2)-y(1 \!+ \!\left| {{{\bm{v}}^H}{\bar{\bm{g}}_1} } \right|^2)+\ln y\nonumber\\
{\rm s.t.}\quad& y>0 {\;{\rm and}\;} \eqref{eqcons1}.
\end{align}
This reformulated problem is still non-convex, but it is convex in $\bm v$ or $y$ with the other variable is fixed. Then, we can further apply alternating optimizing to optimize ${\bm{v}}$ and $y$ in an iterative manner. Specifically, the alternating optimization subproblems are given as follows:
\begin{align}
({\rm {\bf P}}6\!-\!{\rm A}):
\mathop {\max }\limits_{y>0}\;\;  -y(1 \!+ \!\left| {{{\bm{v}}^H}{\bar{\bm{g}}_1} } \right|^2)+\ln y,\nonumber
\end{align}
which is a convex optimization problem of $y$ for a given $\bm v$, and
\begin{align}
({\rm {\bf P}}6\!-\!{\rm B}):
\mathop {\max }\limits_{{\bm{V}}\succeq0} \;\;\;&\ln( {1 +  {{\bm{\bar h}}_1^H{\bm V}} }{{{\bm{\bar h}}}_1}) - y  {{\bm{\bar g}}_1^H{\bm V}}{{{\bm{\bar g}}}_1} \nonumber\\
{\rm s.t.}\;\;\;&  { {\rm Tr}( {{\bm{e}}_l^H{\bm V}}{{\bm{e}}_l} )} \le1, 1 \le l\le L+1,
\end{align}
which is a convex optimization problem of ${\bm V}={\bm{v}}{{\bm{v}}^{\bm{H}}}$ for a given $y$, where ${\bm e}_l\in{\mathbb R}^{(L+1)\times1}$ is a unit vector with the $l$-th entry being one and other entries being zero.
Note that since the rank of the optimal ${{\bm{V}}}$ of ({\rm {\bf P}}6-{\rm B}) must be one, which will be proved in Appendix \ref{appendix4}, there is no need to add the rank-one constraint for variable $\bm V$.

In the following, we show the optimal solutions to ({\rm {\bf P}}6-{\rm A}) and ({\rm {\bf P}}6-{\rm B}), respectively. Specifically, according to Lemma  \ref{theorem3}, it is straightforward to know the optimal solution to ({\rm {\bf P}}6-{\rm A}) is given by
\begin{align}
y^*=(1 \!+ \!\left| {{{\bm{v}}^H}{\bar{\bm{g}}_1} } \right|^2)^{-1},\label{eqopy1}
\end{align}

Next, since ({\rm {\bf P}}6-{\rm B}) is a convex problem, it is straightforward to know that strong duality holds for ({\rm {\bf P}}6-{\rm B}) \cite{Boyd}. Hence, we can obtain its optimal solution by solving its dual problem. To begin with, the Lagrangian dual of problem ({\rm {\bf P}}6-{\rm B}) can be written as
\begin{align}
({\rm {\bf P}}6\!-\!{\rm C}):
\mathop {\min }\limits_{\bm{\chi }} \;\; {\cal L}\left( {\bm{\chi }} \right) \quad {\rm s.t.} \;\;{\chi _l} \ge 0,1 \le l \le L,
\end{align}
where ${\bm{\chi }} = \left[ {{\chi _1},\cdots,{\chi _{L+ 1}}} \right]^T\in{\mathbb R}^{(L+1)\times1}$ is the dual variable and
\begin{align}
{\cal L}\left( {\bm{\chi }} \right)&=\mathop {\max }\nolimits_{{\bm{V}}\succeq0}\left\{\right.\ln ( {1 +  {{\bm{\bar h}}_1^H{\bm{V}}{{{\bm{\bar h}}}_1} } } ) - y{{\bm{\bar g}}_1^H{\bm{V}}{{{\bm{\bar g}}}_1}} \nonumber  \\
&\quad\quad\quad + \sum\nolimits_{l = 1}^{L + 1} {{\chi _l}( {1 -  {{\bm{e}}_l^H{\bm{V}}{{\bm{e}}_l} } } )}\left.\right\}.\label{eqlag1}
\end{align}
Then, the optimal solution to ({\rm {\bf P}}6-{\rm B}) can be obtained by iteratively solving \eqref{eqlag1} with fixed $\bm \chi$ and updating $\bm {\chi}$ by sub-gradient methods, e.g., the ellipsoid
method. The details for the subgradient methods have been studied in \cite{yu2006dual}, which are omitted here for brevity.  Then, we have the following theorem.
}\begin{theorem}\label{theorem4}
If the optimal ${\bm{\chi }}^*$ is given, the optimal solution for ({\rm {\bf P}}6-{\rm B}) is given by ${\bm V}^*={\bm v}^*({\bm v}^*)^H$ with
\begin{align}
{\bm v}^*=\frac{{\sqrt {\left( {1 - \frac{1}{{{(\varepsilon^*) ^2}}}} \right)^{+}} }}{\varepsilon^* }{\left( {{\rm diag}\left( {\bm{\chi }}^* \right) + y{{{\bm{\bar g}}}_1}{\bm{\bar g}}_1^H} \right)^{ - 1}}{{{\bm{\bar h}}}_1},\label{eqopv1}
\end{align}
where $\varepsilon^*  \!=\! \left\|\! {{{\left( {{\rm diag}( {\bm{\chi }}^* ) \!+\! y{{{\bm{\bar g}}}_1}{\bm{\bar g}}_1^H} \right)}^{ - 0.5}}{\bm{\bar h}}} \right\|$.
\end{theorem}
\begin{IEEEproof}
Please refer to Appendix \ref{appendix4}.
\end{IEEEproof}

The convergence for the above algorithm is also guaranteed, the proof is similar to the proof of theorem \ref{theorem1}, which is omitted here for brevity. Besides, the detailed  steps of the above algorithm are summarized in Algorithm \ref{algorithm2}.
\begin{algorithm}
\caption{Alternating optimization method}\label{algorithm2}
\small
{{
\begin{algorithmic}[1]
 \STATE Initialize ${\bm w}_1^{(0)}$, ${\bm v}^{(0)}$, ${\bm y}^{(0)}$ and $t=0$.
\REPEAT
  \STATE $t \leftarrow t+1$,
 \STATE Set ${\bm v}={\bm v}^{(t-1)}$, then calculate ${\bm W}^{(t)}$ by \eqref{eqopw1},
 \STATE Set ${\bm v}={\bm v}^{(t-1)}$ and ${\bm w}_1={\bm w}_1^{(t)}$, then calculate ${y}^{(t)}$ by \eqref{eqopy1},
  \STATE  Initialize ${\bm \chi}={\bm \chi}^{(0)}$ and $i=0$,  set ${\bm w}_1={\bm w}_1^{(t)}$ and ${y}={y}^{(t)}$,
 \REPEAT
  \STATE  $i \leftarrow i+1$, set ${\bm \chi}={\bm \chi}^{(i-1)}$, then calculate ${{\bm v}}^{(t)}$ by \eqref{eqopv1},
  \STATE  Calculate ${\bm \chi}^{(i)}$ according to the ellipsoid method \cite{yu2006dual},
  \UNTIL Convergence.
  \UNTIL Convergence.
\end{algorithmic}
}}
\end{algorithm}

\subsection{Heuristic Algorithm for ({\rm {\bf P}}2) with $K\ge1$ and $N\ge 1$}
In this subsection, we provide the heuristic algorithm for ({\rm {\bf P}}2) with $K\ge1$ and $N\ge1$.
Specifically, when $L \to \infty $, we can assume the received signal from the BS to users can be ignored due to the total powers of received signals are dominated by the signals from the BS and through the IRS to the users with asymptotically large $L$. In addition, according to the Rician channel model introduced in \eqref{channel1} and  \eqref{channel2}, when $\kappa_F\to \infty$ and $\kappa_{h_{r,k}}\to \infty$ and $\kappa_{g_{r,n}}\to \infty$, the channel responses from the BS to the IRS, from the IRS to $B_k$ and from the IRS to $E_n$ are dominated by LoS components since the NLoS components can be practically ignored.
Besides, we also assume all legitimate receivers are in the same directions to the IRS, i.e., $\vartheta_{r,k}=\vartheta_{r,i}$ holds for $1\le i,k\le K$. Hence, the total power of received signals at the legitimate receivers can be given by
\begin{align}
{\left| {{\bm{a}}_L^H\left( {{\vartheta_{r,k}}} \right){\bm{\Theta }}{{\bm{a}}_L}(\vartheta^{{\rm{AoA}}})} \right|^2}{\left\| {\left( {{\bm{a}}_M^H(\vartheta ^{{\rm{AoD}}})} \right)\sum\limits_{k = 1}^K {{{\bm{w}}_k}} } \right\|^2},\label{eqpower1}
\end{align}
Then,  it is straightforward to know the optimal ${\bm \theta}^*$ to maximize the total received signal power in \eqref{eqpower1} is given by
\begin{align}
\theta _l^* \!=\! \exp \left( {j\frac{{( {l\!-\! 1} )2\pi \lambda }}{d}( {\sin {\vartheta_{r,k}}\! -\! \sin \vartheta ^{{\rm{AoA}}}} )} \right),\;1\le l\le L.\label{eqoptimaltheta1}
\end{align}

Substituting \eqref{eqoptimaltheta1} into \eqref{eq7} and \eqref{eq8}, the throughput of $k$-th confidential message at $B_k$ and $E_n$ can be written as
\begin{eqnarray}
{\hat R}_k^{\rm{B}}  =\ln\left(1+ \frac{{{\left| { {{\hat {\bm h}}_k}^H   {\bm w}_k} \right|^2}}}{{\sum\nolimits_{i \ne k}^K {{\left|{{{\hat {\bm h}}_k}^H{\bm w}_i } \right|^2} + {\sigma_k ^2}} }}\right),\label{eq71}\\
{\hat R}_{k,n}^{\rm{E}}=\ln\left(1+ \frac{{{{\left|{{\hat {\bm g}}_n}^H{\bm w}_k \right|}^2}}}{{\sum\nolimits_{i \ne k}^K {{{\left| {{\hat {\bm g}}_n}^H{\bm w}_i \right|}^2} + {\delta_n ^2}} }}\right),\label{eq81}
\end{eqnarray}
respectively, where ${{\hat {\bm h}}_k}^H=({{\bm{h}}_{r,k}^H{\bm \Theta} {\bm{F}} + {\bm{h}}_{d,k}^H})$ and ${{\hat {\bm g}}_n}^H=({{\bm{g}}_{r,n}^H{\bm \Theta} {\bm{F}} + {\bm{g}}_{d,n}^H})$.
Next, we apply the ZF beamforming scheme{\footnote{Note that $M\ge N$ is required in the studied system if ZF beamforming is adopted.}}, which forces the information leakage to $E_n$ to be zero, i.e.,
\begin{align}
{\bm {\hat G}}^H{\bm W} ={\bm 0},
\end{align}
where ${\bm{\hat G}} = \left[ {{{{\bm{\hat g}}}_1},\cdots,{{{\bm{\hat g}}}_N}} \right] \in {{\mathbb C}^{M \times N}}$. Thus, the ZF beamformer ${\bm W}$ can be expressed as
\begin{align}
{\bm W}={\bm{X}}{\bm{U}}{\bm P},
\end{align}
where ${\bm X}\in {\mathbb C}^{M\times(M-N)}$ consists of $(M-N)$ singular vectors of $\bm {\hat G}$ corresponding to the zero singular values, ${\bm U}=\left[ {{{\bm u}_1},\cdots,{{\bm{u}}_K}} \right]\in {\mathbb C}^{(M-N)\times K}$ subject to $\left\| {{{\bm u}_k}} \right\|=1$ for $1\le k \le K$. ${\bm P}={\rm diag}({\bm p})\in{\mathbb R}^{K\times K}$ with ${\bm p}=\left[ {{{{p}}_1},\cdots,{p_K}} \right]^T\in{\mathbb R}^{K\times 1}$ subject to $\sum\nolimits_{k = 1}^K {{p_k}}  = P $.

Therefore, problem ({\rm {\bf P}}2) can be changed as the following the minimal-SINR maximization problem, i.e.,
\begin{subequations}
\begin{align}
({\rm {\bf P}}7):\mathop {\max }\limits_{{\bm{U}},{\bm p}} \;\mathop {\min }\limits_{\forall k}\quad &{\gamma_k} \buildrel \Delta \over =  \frac{{{p_k\left| { {{\hat {\bm x}}_k}^H   {\bm u}_k} \right|^2}}}{{\sum\nolimits_{i \ne k}^K {{p_i\left|{{{\hat {\bm x}}_k}^H{\bm u}_i } \right|^2} + 1} }}\nonumber\\
{\rm s.t.}\quad&\sum\nolimits_{k = 1}^K {{p_k}}  = P {\;{\rm and}\;} \left\| {{{\bm u}_k}} \right\|=1.\nonumber
\end{align}
\end{subequations}
where ${\hat {\bm x}}_k^H={\hat {\bm h}}_k^H{\bm X}/{\sigma_k}$. From \cite{cai2011unified}, we know the optimal beamformer in problem ({\rm {\bf P}}2) has the following structure
\begin{align}
{\bm u}_k^*=\frac{{{{\left( {\sum\nolimits_{i \ne k}^K {{z_i{{\bm{\hat x}}}_i}{\bm{\hat x}}_i^H + {{\bm I}_K}} } \right)}^{ - 1}}{{{\bm{\hat x}}}_k}}}{{\left\| {{{\left( {\sum\nolimits_{i \ne k}^K {z_i{{{\bm{\hat x}}}_i}{\bm{\hat x}}_i^H + {{\bm I}_K}} } \right)}^{ - 1}}{{{\bm{\hat x}}}_k}} \right\|}},
\end{align}
where the unique and positive $z_k$\footnote{In fact, $z_k$ is the power allocation for the virtual dual uplink network \cite{cai2011unified}, which is strictly positive} for $1\le k \le K$ can be obtained by solving the following equations \cite{cai2011unified}:
\begin{align}
z_k^*& = \frac{{{\gamma ^*}}}{{{\bm{\hat x}}_k^H{{\left( {\sum\nolimits_{i \ne k}^K {{{{\bm{\hat x}}}_i}{\bm{\hat x}}_i^H + {{\bm I}_K}} } \right)}^{ - 1}}{{{\bm{\hat x}}}_k}}}, \\
{\gamma ^*}& = \frac{P}{{\sum\nolimits_{k = 1}^K \left({{\bm{\hat x}}_k^H{{( {\sum\nolimits_{i \ne k}^K {{{{\bm{\hat x}}}_i}{\bm{\hat x}}_i^H + {{\bm I}_K}} } )}^{ - 1}}{\bm{\hat x}}_k} \right)^{-1}}},
\end{align}
Note that ${\gamma ^*}$ is the optimal value of the objective function in ({\rm {\bf P}}7), and we know ${\gamma_k^*}={\gamma ^*}$ holds for all $1\le k\le K$ \cite{schubert2004solution}. Using this fact, we have the following equation:
\begin{align}
\frac{{\bm{p}}^*}{{{\gamma ^*}}} = {\bm{\Xi Yp^*}} + {\bm{\Xi }}{{\bm{1}}_K},
\end{align}
where ${\bm{\Xi }}={\rm diag}([\frac{1}{{{{| {{\bm{\hat x}}_1^H{{\bm u}_1}} |}^2}}},\cdots,\frac{1}{{{{| {{\bm{\hat x}}_K^H{{\bm u}_K}} |}^2}}}])\in{\mathbb R}^{K\times K}$, ${\bm Y}=\left[ {{Y_{ik}}} \right]\in{\mathbb R}^{K\times K}$ with ${{Y_{ik}}}={| {{\bm{\hat x}}_i^H{{\bm u}_k}} |}^2$ if $i\ne k$ and ${{Y_{ik}}}=0$ if $i = k$. Then, we know the optimal power allocation $\bm p$ is given as follows£º
\begin{align}
{\bm p}^*= {\gamma ^*}\left( {{{\bm{I}}_k} - {\gamma ^*}{\bm{\Xi Y}}} \right){\bm{\Xi }}{{\bm{1}}_K}.
\end{align}

\section{Simulation Results}\label{secSimulation}

In this section, we present simulation results to validate the advantages of using the IRS to improve the secret communication of the downlink MISO broadcast system. It is assumed that the noise variances $\sigma_k^2$ at $B_k$ and $\delta_n^2$ at $E_n$ are the same and normalized to one. The maximum transmit power $P$ is defined in dB with respect to the noise variance. For performance comparison, we also show the performances of two suboptimal baseline schemes. In particular, for the ``Rand" baseline, we randomly select the reflecting coefficients of the IRS from ${\bm {\Phi}}_2 $ with equal probability and apply Algorithm 1 without updating the reflecting coefficients anymore to obtain the beamforming design; and for the  ``Without IRS" baseline, we assume that the channels from the BS to the IRS, from the IRS to $B_k$, and from the IRS to $E_n$ are blocked, i.e., ${\bm F}= {\bm 0}$, ${\bm h}_{r,k}^H={\bm 0}$, and ${\bm g}_{r,n}^H={\bm 0}$, respectively, with $1\le k\le K$ and $1\le n\le N$, and then we apply Algorithm 1 to obtain the beamforming design. This represents the worst-case scenario to achieve secret communication in the absence of the IRS. In addition, we assume that the legitimate receivers and the eavesdroppers are in the same directions to the BS and the IRS, i.e., ${{\vartheta _{d,k}}}={{\vartheta _{d,i}}}$ and ${{\vartheta _{r,k}}}={{\vartheta _{r,i}}}$ for $1\le i,k\le K$, and  ${{\tilde \vartheta _{d,n}}}={{\tilde\vartheta _{d,i}}}$ and ${{\tilde\vartheta _{r,n}}}={{\tilde\vartheta _{r,i}}}$ for $1\le i,n\le N$. We also assume that  ${{\vartheta _{r,k}}}$, ${{\tilde\vartheta _{r,n}}}$, ${\vartheta^{\rm  AoA}}$ and ${\vartheta^{\rm  AoD}}$ are uniformly distributed between $\left[ {0,2\pi } \right)$, while ${{\vartheta _{d,k}}}$ and ${{\tilde \vartheta _{d,n}}}$ are uniformly distributed between $\left[ {-\pi/3,\pi/3 } \right]$.

\begin{figure}
 \centering{
    \includegraphics[width = 0.45\textwidth]{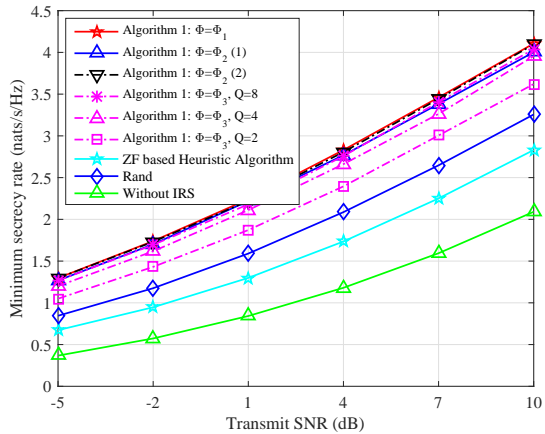}\label{figure2}\subcaption{}
    \includegraphics[width = 0.45\textwidth]{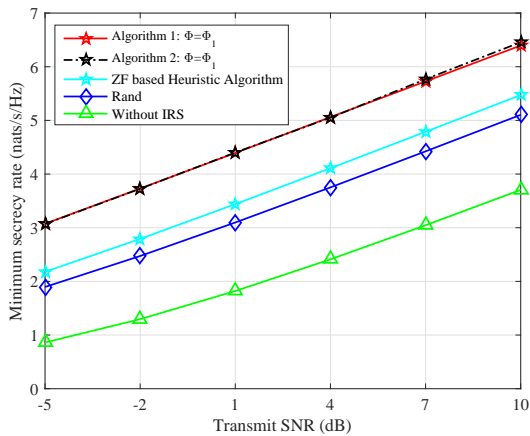}\label{figure21}\subcaption{}
    }
    \caption{Minimum secrecy rate versus the maximum transmit power of the BS: $M = L = 5$, and $\kappa_{\bm h} = 1$ with ${\bm h }\in {\bm {\cal H}}$: (a) $K=N=2$; (b)  $K=N=1$.}
   \label{fig1}
\end{figure}

%
%

Figure \ref{fig1}(a) and Fig. \ref{fig1}(b) show the minimum secrecy rate under different values of the maximum transmit power $P$ of the BS for $K=N=2$ and $K=N=1$, respectively. Note that ``Algorithm 1: ${\bm \Phi_2} (1)$" and ``Algorithm 1: ${\bm \Phi_2} (2)$" denote Algorithm 1 with the first method based on Taylor series expansion and second method based on projection for ${\bm \Phi}={\bm \Phi}_2$, respectively, the details of which are given in subsection \ref{subsecP2}. From the two figures,  we can first observe that the minimum secrecy rates of all methods increase as the maximum transmit power increases. Besides, we know the performance gap between the system with the IRS and the system without the IRS increases with the transmit power, which validates the advantages of the introduced IRS.
Secondly, in the scenario of ${\bm \Phi}={\bm \Phi}_1$, we observe that Algorithm 1 and Algorithm 2 have the similar performances and that both of them outperform the other baselines. This is because both of them can guarantee to converge to a local (global) optimum. Thirdly, we observe that the performance of the ZF based heuristic algorithm is worse than that of ``Rand" baseline with $K=2$ and better than that with $K=1$. This is acceptable due to the complexity of the ``Rand" baseline is still higher than the ZF based heuristic algorithm. It also shows that  the heuristic algorithm is more effective when $K$ is small.

\begin{figure}[t]
\centering
\includegraphics[width=8cm]{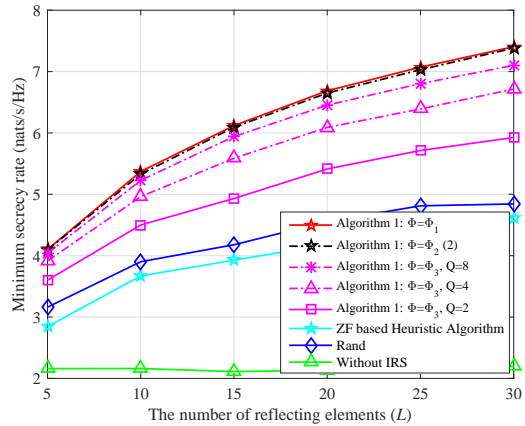}
\caption{Minimum secrecy rate versus the number of reflecting elements on the IRS: $M = 5$, $K=N=2$, $P=10$ dB and $\kappa_{\bm h} = 1$ with ${\bm h }\in {\bm {\cal H}}$.  } \label{figure3}\vspace{-0.3cm}
\end{figure}

Figure \ref{figure3} plots the minimum secrecy rate versus the number of reflecting elements (antennas) $L$ of the IRS. From this figure, we can observe that the minimum secrecy rates of all methods assisted by the IRS increase as the number of reflecting elements on the IRS increases, while the minimum secrecy rate of the system without the IRS remains constant. This is reasonable since a larger number of reflecting elements of the IRS can achieve higher array gain. This also validates the advantages of the introduced IRS for the studied systems. In addition, the performance gap between Algorithm 1 with ${\bm \Phi}={\bm \Phi}_1$ and ${\bm \Phi}={\bm \Phi}_3$ increases as $Q$ decreases, especially for a large $L$. This is because the system with a large $L$ requires a large $Q$ to achieve more precise adjustment for the reflecting coefficients on the IRS. This indicates that in order to better achieve the array again brought by a larger $L$, we should use a larger $Q$ for the proposed scheme with ${\bm \Phi}={\bm \Phi}_3$.


\begin{figure}[t]
\centering
\includegraphics[width=8cm]{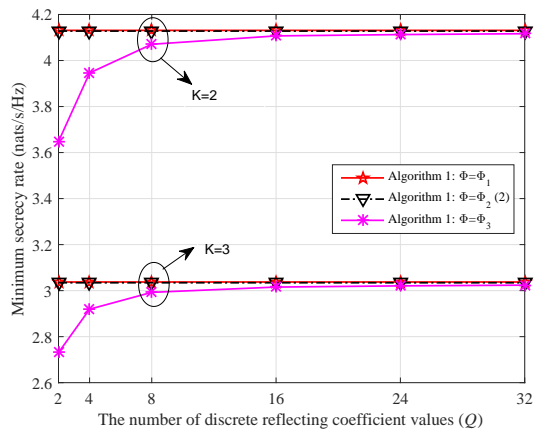}
\caption{Minimum secrecy rate versus the number of reflecting coefficients of each reflecting elements of the IRS: $M =L=5$, $N=2$, $P=10$ dB and $\kappa_{\bm h} = 1$ with ${\bm h }\in {\bm {\cal H}}$.  } \label{figure4}\vspace{-0.3cm}
\end{figure}

Figure \ref{figure4} shows the minimum secrecy rate versus the number of discrete reflecting coefficient values $Q$ of the reflecting coefficients on the IRS. We can observe that the performances of the proposed algorithms with ${\bm \Phi}={\bm \Phi}_1$, ${\bm \Phi}={\bm \Phi}_2$ and ${\bm \Phi}={\bm \Phi}_3$ decrease successively due to the fact  ${\bm \Phi}_3 \subseteq {\bm \Phi}_2  \subseteq  {\bm \Phi}_1$. Furthermore, the performance of the proposed algorithm with ${\bm \Phi}={\bm \Phi}_3$ increases as $Q$ increases.
This is because a larger $Q$ allows a much finer adjustment to the reflecting coefficients on the IRS. Thus, the minimum secrecy rate can be improved. Finally, we observe that the proposed algorithm with ${\bm \Phi}={\bm \Phi}_3$ and $Q=8$ or $16$ can achieve a similar performance to the proposed algorithm with ${\bm \Phi}={\bm \Phi}_2$. 

\begin{figure}[t]
\centering
\includegraphics[width=8cm]{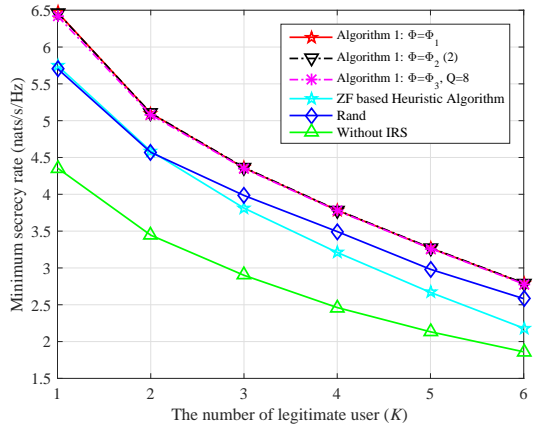}
\caption{Minimum secrecy rate versus the number of legitimate users: $M=10$, $L=5$, $N=2$, $P=10$ dB and $\kappa_{\bm h} = 3$ with ${\bm h }\in {\bm {\cal H}}$.} \label{figureL}\vspace{-0.3cm}
\end{figure}

 Figure \ref{figureL} plots  the minimum secrecy rate versus the number of legitimate receivers $K$. First, we observe that the minimum secrecy rate decrease as $K$ increases, due to the fact that the beamforming gain and array gain need to be shared with more legitimate users. Moreover, it¡¯s similar to Fig. \ref{fig1}, we can observe that the performance of ``Algorithm 1'' is better than that of  ``ZF based Heuristic Algorithm" and other baseline schemes. In addition, we also know that the low-complexity heuristic algorithm is
more effective than the other suboptimal baseline schemes when $K$ is small, especially for $K=1$ or $2$.

\begin{figure}[t]
\centering
\includegraphics[width=8cm]{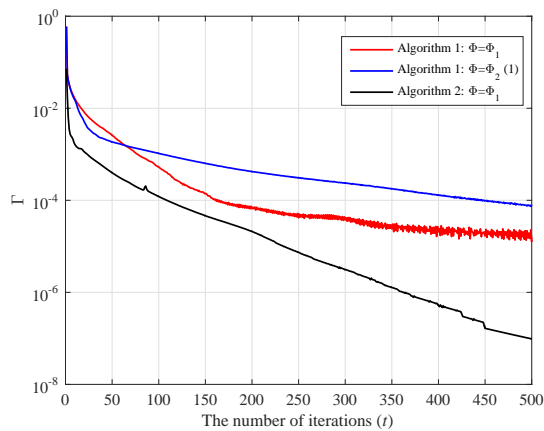}
\caption{Convergence performance versus the number of iterations: $M=5$, $L=5$, $K=1$, $N=1$, $P=5$ dB and $\kappa_{\bm h} = 1$ with ${\bm h }\in {\bm {\cal H}}$.} \label{figure5}\vspace{-0.3cm}
\end{figure}

Figure \ref{figure5} investigates the convergence performances of the proposed algorithms.  For convenience, we first define $\Gamma$ as the normalized performance gap between the values
of the objective function in the two successive iterations of  Algorithm 1 with ${\bm \Phi}={\bm \Phi}_1$, with ${\bm \Phi}={\bm \Phi}_2 (1)$, and Algorithm 2.
We  can first observe that the objective functions in all methods increases with every iteration, which validates the convergence analysis studied in Section \ref{secProblemSolution}.
Moreover, the convergence performance of Algorithm 1 with ${\bm \Phi}={\bm \Phi}_2 (1)$ is worse than that with ${\bm \Phi}={\bm \Phi}_1$. This is because Algorithm 1 with ${\bm \Phi}={\bm \Phi}_2 (1)$ has the main drawback that there is no beforehand choice for the relaxation factor to speed up the convergence \cite{nasir2017secrecy}. Besides, it is worth noting that the number of iterations to achieve convergence in Algorithm 2 is smaller than that in Algorithm 1. This is because in the each iteration of Algorithm 2, we optimize the original problem without approximation and provide the global optimal solution for each subproblem.


\section{Conclusions}\label{secConclusion}
In this paper, we have investigated the joint beamforming and reflecting coefficient designs for a programmable downlink MISO broadcast system with multiple eavesdroppers.
In particular, considering the scenario that the channel responses of the legitimate receivers are highly correlated with those of the eavesdroppers, it is intractable to guarantee the secret communications with the use of beamforming only at the transceivers. Hence, we have explored the use of the IRS to create a programmable wireless environment by providing additional communication links to increase the SNR at the legitimate receivers while suppressing the SNR at the eavesdropper.
Specifically, we have formulated a minimum-secrecy-rate maximization problem under various practical constraints on the reflecting coefficients, which captures the scenarios of both continuous and discrete reflecting coefficients of the reflecting elements. Since the formulated problem is a non-convex problem, we have proposed an efficient algorithm to solve it in an iterative manner and theoretically analyzed its convergence. In addition, we have developed two suboptimal algorithms with closed-form solutions to further reduce the complexity. Finally, the simulation results have validated the advantages of the IRS and the effectiveness of the proposed algorithms.

\appendix

\subsection{Proof of Theorem \ref{seceqapproximationequal}} \label{throem41}
We first prove the lower bound function of $f_{k}^{\rm{B}}\left( {{\bm{W}},{\bm{v}}} \right)$, and then prove the upper bound function of $f_{k,n}^{\rm{E}}\left( {{\bm{W}},{\bm{v}}} \right)$ in this part, which is similar to the proof in \cite{nasir2017secrecy}.

To obtain the lower bound function of $f_{k}^{\rm{B}}\left( {{\bm{W}},{\bm{v}}} \right)$, we first prove the convexity of function $\textstyle{f\left( {x,y} \right) =  - \ln \left( {1 - {{{\left| { x} \right|^2}}}/{y}} \right)}$.
Since $ - \ln \left( 1 - z \right)$ is an increasing and convex function  with respect to $z$ and $z={{{\left| { x} \right|^2}}}/{y}$ is a convex function with respect to $(x,y)$ in the domain $\left\{ {(x,y)\left| {0 \le y \le {\left| { x} \right|^2}} \right.} \right\}$, $f\left( {x,y} \right)$ is thus a convex function. According to the first-order Taylor series expansion of $f\left( {x,y} \right) $ at $({\tilde x},{\tilde y})$, we have
\begin{align}
&f\left( {x,y} \right) \ge f\left( {\tilde x,\tilde y} \right) + {\nabla _{\tilde x}}f\left( {x,\tilde y} \right)\left( {x - \tilde x} \right) + {\nabla _{\tilde y}}f\left( {\tilde x,y} \right)\left( {y - \tilde y} \right)\nonumber\\
& = f\left( {\tilde x,\tilde y} \right) + 2\frac{{\Re \left\{ {\tilde x\left( {x - \tilde x} \right)} \right\}}}{{\tilde y - {\left| {\tilde x} \right|}^2}} - \frac{{{{\left| {\tilde x} \right|}^2}}}{{\tilde y\left( {\tilde y - {{\left| {\tilde x} \right|}^2}} \right)}}\left( {y - \tilde y} \right).
\end{align}
Then, setting $b=y-\left| x \right|^2$ and ${\tilde b}={\tilde y}-{ \left| {\tilde x} \right|}^2$, we have
\begin{align}
\ln ( {1 + \frac{{{{\left| x \right|}^2}}}{b}} ) \ge& \ln ( {1 + \frac{{{{\left| {\tilde x} \right|}^2}}}{{\tilde b}}} ) + 2\frac{{\Re \left\{ {\tilde xx} \right\}}}{{\tilde b}} \nonumber\\
&- \frac{{{{\left| {\tilde x} \right|}^2}}}{{\tilde b( {\tilde b + {{\left| {\tilde x} \right|}^2}} )}}( {b + {{\left| { x} \right|}^2}} ) - \frac{{{{\left| {\tilde x} \right|}^2}}}{{\tilde b}}.
\end{align}
Finally, letting $x={{{{\bm v}^H{\bm H}_k{\bm w}_k }}}$, $b={b_{k}\left( {{\bm{W}},{\bm{v }}} \right)}$, ${\tilde x}= {{{\left( {{{\bm{v}}^{( t )}}} \right)}^H}{{\bm{H}}_k}{\bm{w}}_k^{( t )}}$ and ${\tilde b}={b_{k}( {{\bm{W}}^{(t)},{\bm{v }}^{(t)}} )}$, we can obtain \eqref{path1}.

To obtain the lower bound function of $f_{k}^{\rm{B}}\left( {{\bm{W}},{\bm{v}}} \right)$, since function $\ln(1+z)$ is concave function with respect to $z$, we have
\begin{align}
\ln \left( {1 + z} \right) \le \ln \left( {1 + \bar z} \right) + {{\left( {z - \bar z} \right)}}/({{1 + \bar z}}).
\end{align}
Then, we have
\begin{align}
&f_{k,n}^{\rm{E}}( {{\bm{W}},{\bm{\theta }}} )
\le\textstyle{ f_{k,n}^{\rm{E}}( {{{\bm{W}}^{^{( t )}}},{{\bm{v}}^{^{( t )}}}} )\!  + \! ( {1\!  + \! \frac{{{{\left| {{{( {{{\bm{v}}^{( t )}}} )}^H}{{\bm{G}}_n}{\bm{w}}_k^{( t )}} \right|}^2}}}{{{q_{k,n}}( {{{\bm{W}}^{^{( t )}}},{{\bm{v}}^{^{( t )}}}} )}}})^{ - 1}}\nonumber\\
&\times\textstyle{(  \frac{{{{\left| {{{\bm{v}}^H}{{\bm{G}}_n}{{\bm{w}}_k}} \right|}^2}}}{{{q_{k,n}}( {{\bm{W}},{\bm{v}}} )}} -   \frac{{{{\left| {{{( {{{\bm{v}}^{( t )}}} )}^H}{{\bm{G}}_n}{\bm{w}}_k^{( t )}} \right|}^2}}}{{{q_{k,n}}( {{{\bm{W}}^{^{( t )}}},{{\bm{v}}^{^{( t )}}}} )}}  )}.\nonumber\\
\end{align}
Then, since it is straightforward to know ${{{q_{k,n}}( {{\bm{W}},{\bm{v}}} )}}\ge{{{{q_{k,n}}( {{\bm{W}},{\bm{v}};{{\bm{W}}^{^{( t )}}},{{\bm{v}}^{^{( t )}}}} )}}}$, we have $\frac{{{{\left| {{{\bm{v}}^H}{{\bm{G}}_n}{{\bm{w}}_k}} \right|}^2}}}{{{q_{k,n}}( {{\bm{W}},{\bm{v}}} )}}\le\frac{{{{\left| {{{\bm{v}}^H}{{\bm{G}}_n}{{\bm{w}}_k}} \right|}^2}}}{{{q_{k,n}}( {{\bm{W}},{\bm{v}}};{{\bm{W}}^{t},{\bm{v}}^{t}} )}}$. Finally, we have \eqref{path2}.

\subsection{Proof of Theorem \ref{theorem2}} \label{appendix2}
As aforementioned, the sequence of $({\bm W}^{(t)},{\bm v}^{(t)})$ will converges to $({\bm W}^*,{\bm v}^*)$ as $t\to \infty $. Next, we write the Lagrangian function of $({\rm {\bf P}}2\!-\!t)$ as
\begin{align}
 {\cal L}( {{\bm{W}},{\bm{v}},{\bm{\eta }},\chi } ) =& {{\cal R}^{{\rm{lb}}}}( {{{\bm{W}}^{(t)}},{{\bm{v}}^{(t)}};{{\bm{W}}^{(t)}},{{\bm{v}}^{(t)}}} )\nonumber\\
  &\!\!\!\!\!\!\!\!\!\!\!\!\!\!\!\!\!\!\!\!\!\!\!\!\!+ \sum\nolimits_{l = 1}^{L + 1} {{\eta _l}} ( {1 - {v_l}} ) + \chi ( {P - \sum\nolimits_{k = 1}^K  {\left\| {{{\bf{w}}_k}} \right\|^2} } ),
\end{align}
where  ${\bm{\eta }} = [ {{\eta _1},.,{\eta _{L + 1}}} ]{\in}{\mathbb C}^{1\times{(L+1)}}$ and $\chi $ are the dual variables. Then, when $t\to \infty $, the corresponding KKT conditions are given as follows:
\begin{align}
&{\nabla _{{v_n}}}{{\cal R}^{{\rm{lb}}}}( {{{\bm{W}}^*},{{\bm{v}}^*};{{\bm{W}}^*},{{\bm{v}}^*}} ) - \eta _l^* = 0,\forall l,\\
&{\nabla _{{{\bm{w}}_k}}}{{\cal R}^{{\rm{lb}}}}( {{{\bm{W}}^*},{{\bm{v}}^*};{{\bm{W}}^*},{{\bm{v}}^*}} ) - 2{\chi ^*}{\bm{w}}_k^* = 0,\forall k,\\
&\eta _l^*( {1 - v_l^*} ) = 0,\forall l,\\
&{\chi ^*}( {P - \sum\nolimits_{k = 1}^K {{{\left\| {{\bm{w}}_k^*} \right\|}^2}} } ) = 0.
\end{align}
According to \eqref{eq19}, and  notice that ${\bm W}^{*}$ and ${\bm v}^{*}$ are the optimal solutions of the convex optimization problems of $({\rm {\bf P}}3-{\rm A})$ and $({\rm {\bf P}}3-{\rm B})$, respectively, we know the above KKT conditions are all satisfied. Hence, the converged solution $({\bm W}^{*},{\bm v}^{*})$  is a KKT point.

\subsection{Proof of Theorem \ref{theorem4}} \label{appendix4}
Firstly, \eqref{eqlag1} can be rewritten as follows:
\begin{align}
\!\!\!\!\mathop {\max }\limits_{{\bm{V}}\succeq0}\;\ln ( {1\! +\!  {{\bm{\bar h}}_1^H\!{\bm{V}} {{{\bm{\bar h}}}_1}} } )\!-\!{\rm Tr}(({\rm diag }({\bm \chi})\!+\! y( {{{{\bm{\bar g}}}_1}{\bm{\bar g}}_1^H} )){\bm{V}}).\label{eqa1}
\end{align}
Let us define
\begin{align}
\!\!{ {\bm{\tilde V}}}\!\!=\!\!({\rm diag }({\bm \chi})\!+\! y( {{{{\bm{\bar g}}}_1}{\bm{\bar g}}_1^H} \!))^{0.5}{\bm{V}}({\rm diag }({\bm \chi})\!+\! y( {{{{\bm{\bar g}}}_1}{\bm{\bar g}}_1^H}\! ))^{0.5},
\end{align}
then, \eqref{eqa1} can be rewritten as
\begin{align}
\!\!\!\!\mathop {\max }\nolimits_{{\bm{\tilde V}}\succeq0}\;\ln ( {1\! +\!  \varepsilon^2 {\bm c}^H{{\bm{\tilde V}} } {{{\bm{c}}}}} )\!-\!{\rm Tr}({\bm{\tilde V}}),\label{eqa3}
\end{align}
where
\begin{align}
{\bm c}&={{1}/{\varepsilon}}{({\rm diag }({\bm \chi})\!+\! y( {{{{\bm{\bar g}}}_1}{\bm{\bar g}}_1^H} \!))^{-0.5}{\bm{\bar h}}_1 },  \\
\varepsilon&={{\left\| \!({\rm diag }({\bm \chi})\!+\! y( {{{{\bm{\bar g}}}_1}{\bm{\bar g}}_1^H} \!))^{-0.5}{\bm{\bar h}}_1 \right\|}}.
\end{align}
It is straightforward to know that the rank of the optimal ${\bm{\tilde V}}$ must be one \cite{zhang2010cooperative}, which can be represented as ${\bm{\tilde V}}=w{\bm{\tilde v}}{\bm{\tilde v}}^H$ where ${\left\| {\bm{\tilde v}} \right\|^2}=1$. Then, from \eqref{eqa3},   the optimal ${\bm{\tilde v}}^*$ is ${\bm c}$, and $w$ can be obtained by solve the following problem:
\begin{align}
\!\!\!\!\mathop {\max }\nolimits_{w\ge0}\;\ln ( {1\! +\!  \varepsilon^2 w })- w.\label{eqa4}
\end{align}
Obviously, the optimal $w^*$ is $\left( {1 - \frac{1}{{{\varepsilon ^2}}}} \right)^{+}$. Thus, we have proved this theorem.

\ifCLASSOPTIONcaptionsoff
  \newpage
\fi


\end{document}